\documentclass[12pt]{article}

\usepackage{lineno,hyperref}
\usepackage[ruled,linesnumbered, vlined]{algorithm2e}
\usepackage[noend]{algpseudocode}
\usepackage{multirow}
\usepackage{multicol}
\usepackage{booktabs} % For formal tables
\usepackage{adjustbox}
\usepackage{threeparttable}
\usepackage{authblk}

\usepackage{amsmath,amssymb,amsfonts}
\usepackage{algorithmicx}
\usepackage{graphicx}
\usepackage{textcomp}
\usepackage[dvipsnames]{xcolor}
\def\BibTeX{{\rm B\kern-.05em{\sc i\kern-.025em b}\kern-.08em
		T\kern-.1667em\lower.7ex\hbox{E}\kern-.125emX}}

\definecolor{ForestGreen}{RGB}{63,142,38}
\definecolor{UnibasMint}{RGB}{30,165,165}
\definecolor{UnibasGrey}{RGB}{140,145,150}

\definecolor{darkblue}{rgb}{0.1, 0.1, 0.44}

\definecolor{findNeighbors}{HTML}{0000FF}
\definecolor{Density}{HTML}{000075}
\definecolor{IAD}{HTML}{3791FF}
\definecolor{Divv}{HTML}{0B60FF}
\definecolor{Momentum}{HTML}{69CBFF}
\definecolor{Gravity}{HTML}{00AA00}
\definecolor{Communication}{HTML}{FF0000}
\definecolor{openMP}{HTML}{FE00FF}

\usepackage[normalem]{ulem}
\newcommand{\ahmed}[1]{{\color{black}#1}}
\newcommand{\alir}[1]{{\color{black}#1}}

\newcommand{\fk}[1]{{\color{black}#1}}
\newcommand{\fmc}[1]{{\color{black}#1}}

\newcommand{\ali}[1]{{\color{black}#1}}

\newcommand{\simdag}{\mbox{SG-SD}}
\newcommand{\msg}{\mbox{SG-MSG}}
\newcommand{\smpi}{\mbox{SG-SMPI}}
\newcommand{\simgrid}{\mbox{SG}}

\newcommand{\dlbTool}{\mbox{\textit{DLB\_tool}}}

\newcommand{\tl}{\mbox{thread-level}}
\newcommand{\pl}{\mbox{process-level}}
\newcommand{\cut}[1]{}

\begin{document}

\title{An Approach for Realistically Simulating the Performance of Scientific Applications on High Performance Computing Systems}

%
%\author{Franziska Kasielke}
%\address{Institute of Software Methods for Product Virtualization, German Aerospace Center, Germany}
%\ead{franziska.kasielke@gmx.de}
%
%\author{Ioana Banicescu}
%\address{Department of Computer Science and Engineering, Mississippi State University, USA}
%\ead{ioana@cse.msstate.edu}

\author{Ali Mohammed}
\author{Ahmed Eleliemy}
\author{Florina M. Ciorba}	
\affil{Department of Mathematics and Computer Science\\
	University of Basel, Switzerland\\}

\renewcommand\Authands{ and }

\author{Franziska Kasielke}
\affil{Institute of Software Methods for Product Virtualization, German Aerospace Center, Germany}

\author{Ioana Banicescu}
\affil{Department of Computer Science and Engineering, Mississippi State University, USA}

\maketitle
\clearpage

\tableofcontents
\clearpage

% !TEX root =  fgcs_19.tex
\begin{abstract}
Scientific applications often contain large, \mbox{computationally-intensive}, and irregular parallel loops or tasks that exhibit stochastic characteristics. 
Applications may suffer from load imbalance during their execution on high-performance computing~(HPC) systems due to such characteristics. 
Dynamic loop \mbox{self-scheduling}~(DLS) techniques are instrumental in improving the performance of scientific applications on HPC systems via load balancing. 
Selecting a DLS technique that results in the best performance for different problems and system sizes requires a large number of exploratory experiments. 
A theoretical model that can be used to predict the scheduling technique that yields the best performance for a given problem and system has not yet been identified. 
Therefore, simulation is the most appropriate approach for conducting such exploratory experiments with reasonable costs. 
This work devises an approach to realistically simulate \mbox{computationally-intensive} scientific applications that employ DLS and execute on HPC systems. 
Several approaches to represent the application tasks (or loop iterations) are compared to establish their influence on the simulative application performance. 
A novel simulation strategy is introduced, which transforms a native application code into a simulative code. 
The native and simulative performance of two \mbox{computationally-intensive} scientific applications are compared to evaluate the realism of the proposed simulation approach. 
The comparison of the performance characteristics extracted from the native and simulative performance shows that the proposed simulation approach fully captured most of the performance characteristics of interest. 
This work shows and establishes the importance of simulations that realistically predict the performance of DLS techniques for different applications and system configurations.
	
\end{abstract}

\textbf{keywords.}
	High performance computing, Scientific applications, \mbox{Self-scheduling}, Dynamic load balancing, Modeling and Simulation, Modeling and simulation of HPC systems, HPC Benchmarking

% !TEX root =  fgcs_19.tex
\section{Introduction}
\label{sec:intro}

%\begin{enumerate}
%	\item a general overview of the field
%	\item motivation
%	\item problem statement
%	\item existing solutions 
%	\item the new solution - high-level illustration
%	\item assumptions and limitations, follow a + - + pattern, i.e., start with positive enthusiastic comments about new work and the contribution, then become realistic and list all the drawbacks and limitations, but then finish on a positive note, with a clear statement about the value of the new contribution
%	\item analysis and simulation
%	\item comparison with the best competing solutions
%	\item contributions
%	\begin{enumerate}
%		\item Novel approach to represent applications and computing systems in simulation
%		\item  Inclusion of advanced adaptive dynamic loop balancing (DLS) techniques
%		\item Two additional scientific applications
%	\end{enumerate}
%	\item paper structure
%\end{enumerate}

Scientific applications are complex, large, and contain irregular parallel loops (or tasks) that often exhibit stochastic behavior. 
The use of efficient loop scheduling techniques, from fully static to fully dynamic, in computationally-intensive applications is crucial for improving their performance on high performance computing (HPC) systems often degraded by load imbalance.
Dynamic loop self-scheduling (DLS) is an effective scheduling approach employed \alir{to improve} \mbox{computationally-intensive} scientific applications \alir{performance} via dynamic load balancing. 
The goal of using DLS is to optimize the performance of scientific applications in the presence of load imbalance caused by \emph{problem}, \emph{algorithmic}, and \emph{systemic} characteristics.
HPC systems become larger on the road to Exascale \alir{computing}.
\alir{Therefore,} scheduling and load balancing become crucial as increasing the number of PEs \alir{leads to} increase \alir{in} load imbalance \fmc{and, consequently, to loss in performance.}

Scheduling and load balancing, from operating system level to HPC batch scheduling level, \alir{in addition to} minimizing the management overhead, are among the most important challenges on the road to Exascale systems~\cite{bergman2008exascale}.  
The static and dynamic loop self-scheduling (DLS) techniques play an essential role in improving the performance of scientific applications. 
These techniques balance the assignment and the execution of independent tasks or loop iterations across the available processing elements (PEs).
Identifying the \fmc{best scheduling strategy} among the available DLS techniques for a given application requires intensive assessment and a large number of exploratory native experiments.
This significant amount of experiments may not always be feasible or practical, due to their associated time and costs.
Simulation mitigates such costs and, therefore, \alir{it} has been shown to be more appropriate \alir{for studying and improving} the \fmc{performance of scientific applications}~\cite{stanisic2015faithful}.
An important source of uncertainty in the performance results obtained via simulation is the degree of trustworthiness in the simulation, understood as the close quantitative and qualitative agreement with the native measured performance. 
Attaining a high degree of trustworthiness eliminates such uncertainty for \fmc{present and future more complex experiments}.

Simulation allows the study of application performance in controlled and reproducible environments~\cite{stanisic2015faithful}.
%Realistic simulation predictions lead to improved application performance in native executions.
Realistic predictions based on trustworthy simulations can be used to design targeted native experiments with the ultimate goal of achieving optimized application performance.
Realistically simulating application performance is, however, nontrivial.
Several studies addressed the topic of application performance simulation for specific purposes, such as evaluating the performance of scheduling techniques under variable task execution times with a specific runtime system~\cite{beaumont2018influence}, or focusing on improving communications in large and distributed applications~\cite{source2sourceISC}.

The present work gathers the authors' \mbox{in-depth} expertise in simulating scientific applications' performance to enable research studies \alir{on} the effects and benefits of employing dynamic load balancing in \mbox{computationally-intensive} applications via \mbox{self-scheduling}.
Several details of representing the application and the computing system characteristics in the simulation are presented and discussed, such as capturing the variability of native execution performance over multiple repetitions as well as calibrating and \mbox{fine-tuning} the simulated system representation for the execution of a specific application.
\fmc{The coupling between the application and the computing system representation has been shown to yield a very close agreement between the native and the simulative experimental results, and to achieve realistic simulative performance predictions~\cite{Mohammed:2018a}.}

\begin{figure}[!h]
	\centering
	\includegraphics[clip, trim=0cm 4cm 15cm 0cm, scale=0.65]{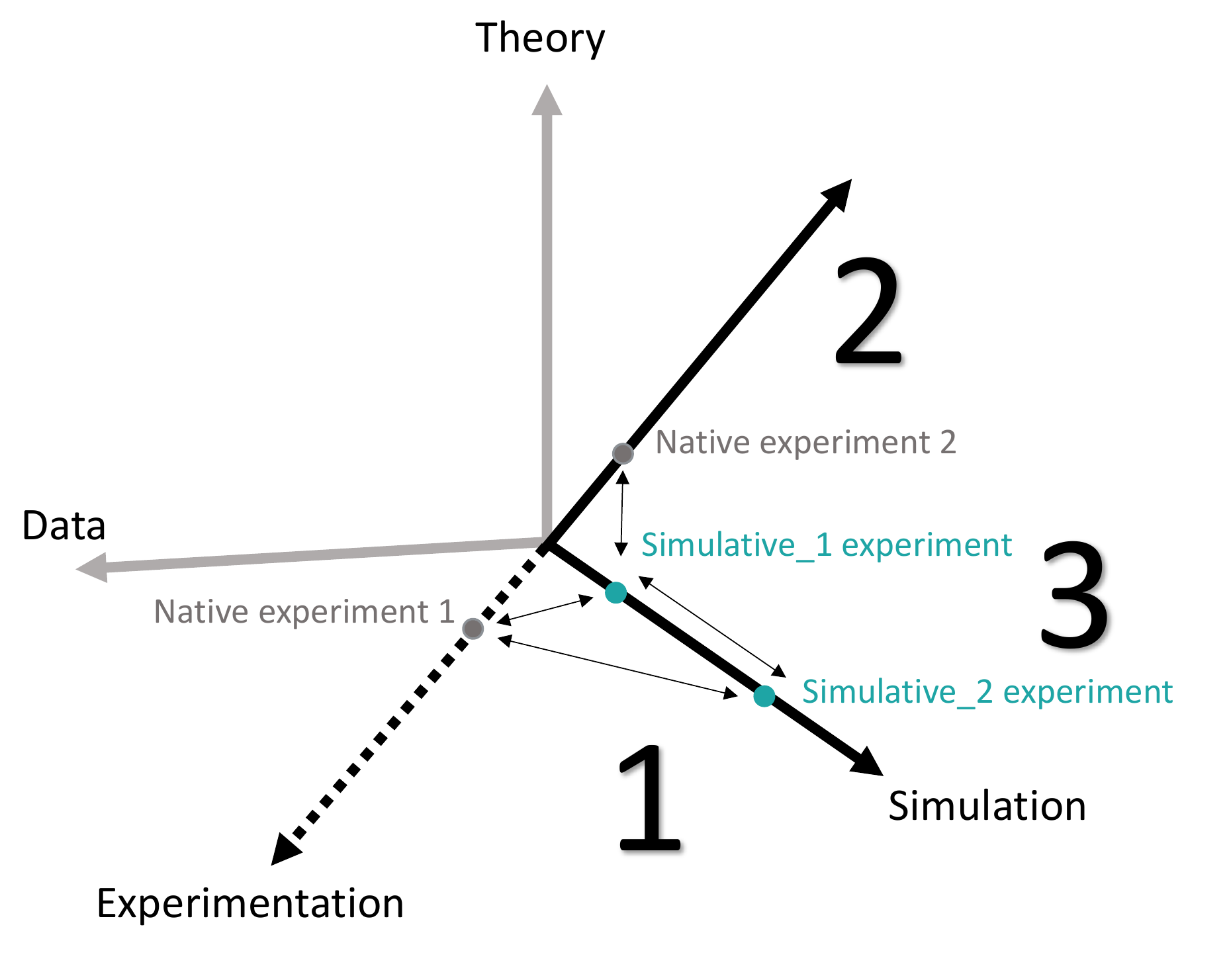}%
	\caption{Illustration of the comparison approach, illustrated over the pillars of science employed in this work for the verification of the DLS techniques: (1)~Native experiments from the past original work from the literature are reproduced in present simulations to verify DLS techniques implementation. (2)~Simulative and native results from experiments in the present are compared to verify the trustworthy simulation of application performance. (3)~Different simulation approaches are compared to achieve close agreement in terms of simulation of application performance  to that of the native performance.}
	\label{fig:comp_approach}
\end{figure}

The proposed \fmc{\emph{realistic simulation}} \fk{approach} is built upon three perspectives of comparison of the results of 
native and simulative experiments, which are also illustrated in \figurename{~\ref{fig:comp_approach}}:
\begin{description}
\item[(1)] \mbox{native-to-simulative} \fk{(or the past)},
\item[(2)] \mbox{native-to-simulative} \fk{(or the present)}, and
\item[(3)] \mbox{simulative-to-native}.
\end{description}

Through the first perspective, the performance reported in \alir{the} original publications, \alir{which} introduced the most \mbox{well-known}, successful, and currently used DLS techniques from the past, is \alir{presently} reproduced via simulation to verify the the similarity in performance results \alir{between} the current DLS techniques implementations and their original implementation~\cite{HPCS}.

In the second perspective, the performance of the \alir{present} native scheduling experiments on HPC systems is compared against that of the simulative experiments. 
This comparison typically enables one to verify and justify the level of the agreement between the results of the native and the simulative experiments, and to answer the question of \emph{``\alir{H}ow realistic are the simulations of applications performance on HPC systems\alir{?''}}~\cite{Mohammed:2018a}.

In the third comparison perspective, different representations of the same application or of the computing system characteristics are used in different simulations.
The simulative performance of the application obtained when employing different DLS technique is compared among different simulative experiments. 
Given that different simulations are expected to represent the same application and platform characteristics, this comparison allows a better assessment of the influence of application and/or system representation on the obtained simulative performance and the degree of agreement between the native and the simulative performance.

The present work makes the following contributions:
(1)~A\fk{n approach} for simulating application performance with a high degree of trustworthiness while considering different sources of variability in application and computing system representations.
(2) A novel simulation \fmc{strategy} of \mbox{computationally-intensive} applications by combining two interfaces of \mbox{SimGrid}~\cite{simgrid} simulation toolkit (SMPI and MSG) to achieve fast and accurate performance simulation with minimal code changes to the native application.
(3)~A realistic simulation of the performance of two scientific applications with several dynamic load balancing techniques.  
\alir{The} applications performance is analyzed based on native and simulative performance results. The performance comparison shows that simulations realistically captured key applications performance features. 
(4)~\alir{An} experimental verification and validation of the use of the different \mbox{SimGrid} interfaces \alir{for representing} the application's tasks characteristics to develop and test DLS techniques in the simulation.

The present work builds upon and extends own prior work~\cite{Mohammed:2018a}~\cite{HPCS}, which focused on the experimental verification of DLS implementation via reproduction~\cite{HPCS} and the experimental verification of application's performance simulation on HPC systems~\cite{Mohammed:2018a}, respectively. 
In the present work, a new \fmc{method} to represent the computational effort in tasks is explored and tested~(c.f. Section~\ref{subsec:method_app}). 
Methods to evaluate and represent variability in the system are also considered in the present work~(c.f. Section~\ref{subsec:method_system}).
An additional scientific application is also included herein~(c.f. Section~\ref{sec:ex}).
The performance of the two scientific applications is examined with four additional adaptive DLS techniques and four additional nonadaptive \alir{DLS} techniques by employing an \mbox{MPI-based} load balancing library both, in native and simulative experiments~(c.f. Section~\ref{subsec:method_dls}).
\fmc{A~novel strategy for simulating applications} is also experimented in this work~(c.f. Section~\ref{subsec:realsim}).
A full version of this manuscript is under publication in the Future Generations Computer Systems Journal, ``On The Road to Exascale II Special Issue: Advances in High Performance Computing and Simulations''.

The remainder of this manuscript is structured as follows. 
Section~\ref{sec:background} presents the relevant background on dynamic load balancing via \mbox{self-scheduling} and the used simulation toolkit.
Section~\ref{sec:related_work} reviews \alir{recent} related work and the \alir{various} simulation approaches adopted therein.
The proposed simulation \fk{approach} is introduced and discussed in Section~\ref{sec:methodology}.
The design of the evaluation experiments, the practical steps of representing the scientific applications in simulation, the results of the native and simulative \alir{experimental} results with various DLS techniques, \alir{as well as their comparisons} are discussed in Section~\ref{sec:ex}. 
Section~\ref{sec:conc} \alir{presents conclusions and an outline of the work envisioned for the future.}

% !TEX root =  fgcs_19.tex
\section{Background}
\label{sec:background}
%Scientific applications are often large and complex.
%Parallel \mbox{computationally-intensive} loops in scientific applications are considered vast resource of independent parallel tasks.
%The performance of scientific applications on HPC systems may suffer due to load imbalance arising from irregular application's or system's characteristics.
%Dynamic loop \mbox{self-scheduling}~(DLS) is often employed to achieve a load balanced execution of scientific applications and improve their performance on large HPC systems.

This section presents and organizes the relevant background of the present work in three dimensions.
The first dimension covers the relevant information concerning dynamic load balancing via dynamic loop \mbox{self-scheduling} techniques, specifically, the selected DLS techniques of the present work.
The second dimension discusses specific research efforts from the literature where DLS techniques enhanced the performance of various scientific applications.
The last dimension introduces the simulation toolkit used in the present work.

%\begin{itemize}
%	\item DLS techniques
%	\item DLS in scientific applications
%	\item SimGrid
%	\item DLS in Simulation - cite Franziska's ISPDC 17 paper \\
%\end{itemize}

\noindent\textbf{Dynamic load balancing via dynamic loop \mbox{self-scheduling}.}
There are two main categories of loop scheduling techniques: static and dynamic. 
The essential difference between static and dynamic loop scheduling is the time when the scheduling decisions are taken.
Static scheduling techniques, such as block, cyclic, and block-cyclic~\cite{li1993locality}, divide and assign the loop iterations (or tasks) across the processing elements~(PEs) before the application executes. 
The task division and assignment do not change during execution.
In the present work, block scheduling is considered and is denoted as STATIC.

Dynamic loop \mbox{self-scheduling} (DLS) techniques divide and \emph{\mbox{self-schedule}} the tasks during the execution of the application. 
As a result, DLS techniques balance the execution of the loop iterations at the cost of increased overhead compared to the static techniques.
\mbox{Self-scheduling} differs from \emph{work sharing}, another related scheduling approach, wherein tasks are assigned onto PEs in predetermined sizes and order~\cite{blumofe1998space}.
Self-scheduling is also different from \emph{work~stealing}~\cite{blumofe1999scheduling} in that PEs request work from a central work queue as opposed to distributed work queues.
The former has the advantage of global scheduling information while the latter is more scalable at the cost of identifying overloaded PEs from which to steal work.
DLS techniques consider independent tasks or loop iterations of applications~\cite{SS,FSC,GSS,FAC,AWF,AWFBC}.
For dependent tasks, several loop transformations, such as loop peeling, loop fission, loop fusion, and loop unrolling can be used to eliminate loop dependencies~\cite{surveyloop}.
DLS techniques can be categorized as \emph{nonadaptive} and \emph{adaptive}~\cite{Banicescu_ETNA}. 
During the application execution, the \mbox{nonadaptive} techniques calculate the number of iterations comprising a chunk based on certain parameters that can be obtained prior to the application execution.
The nonadaptive DLS techniques considered in this work include: modified \mbox{fixed-size} chunk~\cite{banicescu:2013a}~(mFSC), guided self-scheduling~\cite{GSS}~(GSS), and factoring~\cite{FAC}~(FAC).

mFSC~\cite{banicescu:2013a} groups iterations into chunks at each scheduling round to avoid the large overhead of single loop iterations being assigned at a time.  
In mFSC, the chunk size is fixed and plays a critical role in determining the performance of this technique. 
mFSC assigns a chunk size that results in a number of chunks that is similar to that of FAC (explained below).

GSS~\cite{GSS} assigns chunks of decreasing sizes to reduce scheduling overhead and improve load balancing. 
Upon a work request, the remaining loop iterations are divided by the total number of PEs.

FAC~\cite{FAC} improves GSS by scheduling the loop iterations in batches of \mbox{equal-sized} chunks. 
The initial chunk size of GSS is usually larger than the size of the initial chunk using FAC.
If more \mbox{time-consuming} loop iterations are at the beginning of the loop, FAC balances the execution better than GSS. 
The chunk calculation in FAC is based on probabilistic analyses to balance the load among the processes, depending on the prior knowledge of the mean $\mu$ and the standard deviation $\sigma$ of the loop iterations execution times.
Since loop characteristics are not known \fmc{a priori} and typical loop characteristics that can cover many probability distributions, a practical implementation of FAC was suggested~\cite{FAC} that assigns half of the remaining work in a batch. 
This work considers this practical implementation.
Compared to STATIC and mFSC, GSS and FAC provide better \mbox{trade-offs} between load balancing and scheduling overhead.

The \emph{adaptive} DLS techniques exploit, during execution, the latest information on the state of both the application and the system to predict the next sizes of the chunks of the iterations to be executed.  
In highly irregular environments, the adaptive DLS techniques balance the execution of the loop iterations significantly better than the \mbox{nonadaptive} techniques.
However, the adaptive techniques may result in significant scheduling overhead compared to the \mbox{nonadaptive} techniques and are, therefore, recommended in cases characterized by highly imbalanced execution. The adaptive DLS techniques include adaptive weighted factoring~\cite{AWF}~(AWF) and its variants~\cite{AWFBC}~AWF-B,~AWF-C,~AWF-D, and~AWF-E.

AWF~\cite{AWF} assigns a weight to each PE that represents its computing speed and adapts the relative PE weights during execution according to their performance.
It is designed for \mbox{time-stepping} applications. 
Therefore, it measures the performance of PEs during previous \mbox{time-steps} and updates the PEs relative weights after each \mbox{time-step} to balance the load according to the computing system's present state.

\mbox{AWF-B}~~\cite{AWFBC} relieves the \mbox{time-stepping} requirement to learn the PE weights.
It learns the PE weights from their performance in previous batches instead of \mbox{time-steps}.

\mbox{AWF-C}~\cite{AWFBC} is similar to \mbox{AWF-B}, however, the PE weights are updated after the execution of each chunk, instead of batch.

\mbox{AWF-D}~\cite{AWFBC} is similar to \mbox{AWF-B}, where the scheduling overhead (time taken to assign a chunk of loop iterations) is taken into account in the weight calculation. %, in addition to the performance of a PE \ali{executing} a  chunk of loop iterations.

\mbox{AWF-E}~\cite{AWFBC} is similar to \mbox{AWF-C}, and takes into account also the scheduling overhead, similar to \mbox{AWF-D}.
%AF is also based on FAC.
%However, it measures the performance of PEs to learn \ali{the} $\mu$ and $\sigma$ per PE during execution.
\\
\noindent\textbf{DLS in scientific applications.}
The DLS techniques have been used in several studies to improve the performance of computationally-intensive scientific applications. 
They are mostly used at the \pl{} to balance the load between processes running on different PEs.
For example, AWF~\cite{AWF} and FAC~\cite{FAC} were used to balance a load of a heat conduction application on an unstructured grid~\cite{banicescu2001load}.  
%AF resulted in the best performance among the other techniques due to the irregularity of the application and the irregular and heterogeneous nature of the computing system.
Nonadaptive and adaptive DLS techniques such as \mbox{self-scheduling}\footnote{\fmc{To be distinguished from the principle of receiver-initiated load balancing through self-scheduling.}}~(SS)~\cite{SS}, GSS~\cite{GSS}, FAC~\cite{FAC}, AWF~\cite{AWF}, and its variants, were used over the years to enhance applications, such as simulations of wave packet dynamics, automatic quadrature routines~\cite{AWFBC}, N-Body simulations~\cite{fractiling}, solar map generation~\cite{boulmier2016towards}, an image denoising model, the simulation of a vector functional coefficient autoregressive~(VFCAR) model for multivariate nonlinear time series~\cite{carino2007tool}, and \fmc{a parallel spin-image algorithm from} computer vision~(PSIA)~\cite{Eleliemy:2017b}.

With the increase in processor core counts per compute node, advanced scheduling techniques, such as the class of \mbox{self-scheduling} mentioned earlier, are also needed at the \tl{}.
To this end, the GNU OpenMP runtime library was extended~\cite{Ciorba:2018}~(LaPeSD libGOMP) to support four additional DLS techniques, namely: fixed-size chunk~\cite{FSC}~(FSC), trapezoid \mbox{self-scheduling}~\cite{tzen1993trapezoid}~TSS, FAC, and RANDOM (in terms of chunk sizes) besides the originally OpenMP scheduling techniques: STATIC, Dynamic, and Guided~(equivalent to GSS~\cite{GSS}). 
The extended GNU runtime library that implements DLS was used to schedule loop iterations in computational benchmarks, such as the NAS parallel~\cite{Nas_parallel} and \mbox{RODINIA}~\cite{rodinia} benchmark suites.
\\
\noindent\textbf{The selected simulation toolkit.}
\mbox{SimGrid}~\cite{simgrid} is a scientific simulation framework for the study of the behavior of \mbox{large-scale} distributed computing systems, such as, the Grid, the Cloud, and \mbox{peer-to-peer} (P2P) systems. 
It provides application programming interfaces (APIs) to simulate various distributed computing systems. 
\mbox{SimGrid}~(hereafter, SG) provides four different APIs for different simulation purposes. 
\mbox{MetaSimGrid}~(\fmc{hereafter}, \msg{}) and \mbox{SimDag}~(\fmc{hereafter}, \simdag{}) provide APIs for the simulation of computational problems expressed as independent tasks or task graphs, respectively.

The \fmc{\mbox{SimGrid}-SMPI interface (hereafter, \smpi{})} provides the functionality for the simulation of programs written using the message passing interface~(MPI) and targets developers interested in the simulation and debugging of their parallel MPI codes.

The newly introduced \fmc{\mbox{SimGrid}-S4U interface (hereafter, \mbox{SG-S4U})} currently supports most of the functionality of the \msg{} interface with the purpose of also incorporating the functionality of the \simdag{} interface over time.

The present work proposes  a novel simulation approach of \mbox{computationally-intensive} applications by combining \smpi{} and \msg{} to achieve fast and accurate performance simulation with minimal code changes to the native application.

% !TEX root =  fgcs_19.tex
\section{Related Work}
\label{sec:related_work}
%\discuss{[Franziska and Ahmed: please revise and add more related works you know about and write about them]}\\
\noindent\textbf{Scheduling in simulation.}
The \msg{} and \simdag{} interfaces of \simgrid{} were used to implement various DLS techniques. 
For instance, eight DLS techniques were implemented using the \msg{} interface in the literature~\cite{mahad}: five \mbox{nonadaptive}, SS~\cite{SS}, FSC~\cite{FSC}, GSS~\cite{GSS}, FAC~\cite{FAC}, and weighted factoring~(WF)~\cite{WF}, and three adaptive techniques, adaptive weighted factoring~(\mbox{AWF-B}, \mbox{AWF-C})~\cite{AWFBC}, and adaptive factoring~(AF)~\cite{AF}.

The weak scalability of these DLS techniques was assessed in the presence of certain load imbalance sources (algorithmic and systemic).
The flexibility, understood as the robustness against perturbations in the PE computing speed, of the same DLS techniques implemented using \msg{} was also studied~\cite{nitin}.
Moreover, the resilience, understood as the robustness against PE failure, of these DLS techniques on a heterogeneous computing system was studied using the \msg{} interface~\cite{dlsmsg}.

Another research effort used the \msg{} interface to reproduce certain experiments of DLS techniques~\cite{hoffeins2017examining}.  
Therein, a successful reproduction of the past DLS experiments was presented. 
The results were compared to experiments from the past available in the literature to verify the implementation of the DLS techniques. 
A similar approach of verifying the implementation of certain DLS techniques via reproduction was proposed using the \simdag{} interface~\cite{Ali_HPCC:17}. 
%The present work aims to assess the usefulness of these two \simgrid{} interfaces for achieving realistic simulations of scientific applications scheduled using the DLS implemented in \simgrid{}."

The relation between batch and application level scheduling was studied in simulation~\cite{Eleliemy:2017a}, using Alea~\cite{klusavcek2010alea} for the batch level scheduling and \simdag{} for the application level scheduling. 
The two simulators were connected and used together to simulate the execution of multiple applications with various scheduling techniques at the batch level and the application level.
It was shown that a holistic solution resulted in a better performance than focusing on improving the performance at each level solely.

\simgrid{} was also used for the study of file management in large distributed systems~\cite{chai2019realistic} to improve applications performance.
The effect of \fmc{variability in task execution times} on the makespan of applications scheduled using StarPU\cite{augonnet2011starpu} on heterogeneous CPU/GPU systems was also studied in simulation~\cite{beaumont2018influence}.
The results showed that the dynamic scheduling of StarPU improves the performance even with irregular tasks execution times.

\noindent\textbf{Realistic simulation approaches.}
A combination of simulation and trace replay was used to guide the choice of the scheduling technique and the granularity of problem decomposition for a geophysics application to tune its performance~\cite{Geosim}.
\smpi{} was used to generate a time independent trace (TiT), a special type of execution trace, of the application with the finest problem decomposition.
This trace was then modified to represent different granularities of problem decomposition.
Traces that represent different decompositions were replayed with different scheduling techniques to identify the decomposition granularity and scheduling technique combination that results in improved application performance.
The scheduling techniques were extracted from the Charm++ runtime to be used in the simulation.
However, the process of trace modification to represent different decomposition is complex, limits the number of explored decompositions, and may result in inaccurate simulation results.

The \mbox{compiler-assisted} native application source code transformation to a code skeleton suitable for structural simulation toolkit~\cite{rodrigues2012improvements}~(SST) was introduced~\cite{source2sourceISC}.
Special pragmas need to be inserted in the source code to simulate computations as certain delays, eliminate large unnecessary memory allocations in simulation, and handle global variable correctly.
This approach was focused on the simulation for the study of communications and network in large computing systems. 
Therefore, the variability of task execution times was not considered explicitly.

StarPU~\cite{augonnet2011starpu} was ported to \msg{} for the study of scheduling of tasks graphs on heterogeneous CPU/GPU systems. 
Tasks execution times were estimated based on the average execution time benchmarked by StarPU.
Both average task execution time and generating \mbox{pseudo-random} numbers with the same average as task execution time were explored.
However, depending on time measurements only may not be adequate for \mbox{fine-grained} tasks.
In addition, porting the StarPU runtime to a simulator interface is challenging and requires \fmc{significant} effort.

The Monte-Carlo method~\cite{metropolis1949monte} was used to improve the simulation of workloads in cloud computing~\cite{bertot2018improving}.
To capture the variation in applications execution time in simulation, the variability in cloud computing systems was quantified and added to task execution times as a probability. 
The simulation was repeated $500$~times, each with different seeds to obtain a similar effect of the dynamic native execution on the clouds.
However, the variation in the application execution time has two components: (1)~the variability in a task execution time due to application characteristics or system characteristics such as nonuniform memory access; (2)~the variability that stems from the computing system resources being perturbed by operating system interference, other applications that share resources, or transient malfunctions.
Considering both components of application performance variability is important for obtaining realistic simulation results.

In this work, a \fmc{novel} simulation \fk{approach} is presented that considers the different factors that affect application performance. 
Guidelines are proposed in Section~\ref{sec:methodology} on how to estimate the tasks execution times and the system characteristics. 
Fine tuning the system representation to \fmc{closely} reflect the system performance \fmc{for} the execution of a certain application is essential.
Reducing the differences between native and simulative experiments by using the same scheduling library in both native and simulative experiments ensures the same scheduling behavior \fmc{in both types of experiments}.
A novel simulation \fmc{method} that combines the use of two \fmc{SimGrid interfaces, namely \smpi{} and \msg{}, is introduced in Section~\ref{subsec:realsim}, which enables the simulation of application performance with minimal code changes. }

%\begin{itemize}
%	\item scheduling in simulation papers
%	\item Using Simulation to Evaluate and Tune the Performance of Dynamic Load Balancing of an Over-Decomposed Geophysics Application \cite{Geosim}
%	\item source to source transformation \cite{source2sourceISC}
%	\item Faithful performance prediction of a dynamic task-based runtime system for heterogeneous multi-core architectures~\cite{stanisic2015faithful}
%	\item Improving Cloud Simulation Using the Monte-Carlo Method~\cite{bertot2018improving}
%	\item It is important to underline the need for a clear cut, clear separation line, between existing work and new ideas being presented in the paper [single paragraph]. 
%\end{itemize}

% !TEX root =  fgcs_19.tex
\section{\fk{Approach} For Realistic Simulations} 
\label{sec:methodology}
%\begin{itemize}
%	\item simulation vs realistic simulation - what does realistic mean? how to achieve it? how to verify? discussion
%	\item describe main idea
%	\item  methodology illustration
%\end{itemize}

A realistic performance simulation \fmc{means that conclusions drawn from the simulative performance results are close} to \ali{those} drawn from the native performance results.
The \ali{close} agreement between both conclusions does \fmc{not necessarily mean } a close agreement between native and simulative application execution times.
 %not necessitate the close agreement between native and simulative application execution times.
For the study of dynamic load balancing and task \mbox{self-scheduling}, the performance of different scheduling techniques relative to others \fmc{is expected to} be preserved between native and simulative experiments.
\fmc{Preserving the expected behavior} \ali{suffices} to draw similar conclusions on the performance of DLS techniques between native and simulative experiments. 

\begin{figure}[!h]
	\centering
	\includegraphics[clip, trim=0cm 0cm 0cm 0cm, scale=0.52]{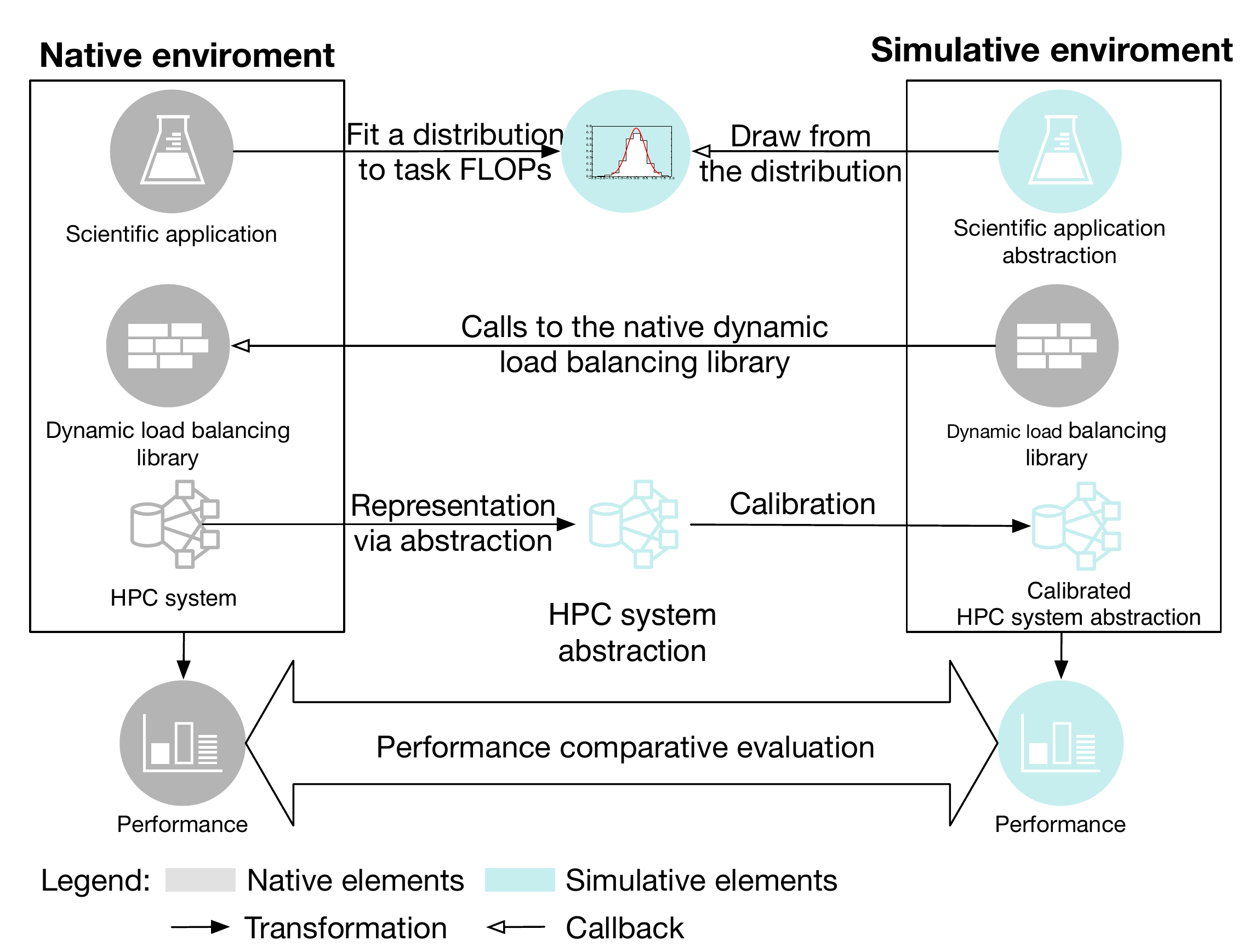}%
	\caption{Illustration of the proposed \ali{generic \fk{approach}} \ali{for realistic simulations}. Scientific application and computing system characteristics \ali{are abstracted for use} in simulation. \ali{A single scheduling library is used which is called both by} the native and simulative executions.}
	\label{fig:sim_approach}
\end{figure}

Preserving \ali{identical} performance characteristics between native and simulation experiments is challenging due to the dynamic interactions between the three main components that affect the performance:
\begin{description}
\item[(1)] Application characteristics,
\item[(2)] Dynamic load balancing, and
\item[(3)] Computing system \fmc{characteristics}.
\end{description}
% (1)~Application characteristics, (2)~Dynamic load balancing, (3)~Computing system. 
\figurename{~\ref{fig:sim_approach}} shows \ali{these} three main performance components and summarizes the proposed simulation approach, where each component \fmc{is independently} represented and verified to achieve realistic simulations.
%Despite the fact that dynamic load balancing of independent tasks in \mbox{computationally-intensive} application is the focus of this work, 
The proposed approach \ali{is generic} \fmc{and} can be adapted \ali{for the systematic and realistic} simulation of other classes of applications, \ali{e.g., \mbox{data-intensive} or \mbox{communication-intensive} applications}, \ali{load balanced using other classes of algorithms}.
The details of representing each component are \ali{provided next}.

\subsection{Representing application\ahmed{s} for realistic simulation\ahmed{s}}	
\label{subsec:method_app}
Two important aspects need to be clear to enable \fmc{the representation of an application in simulation via abstraction}: (1)~ The main application flow, i.e., initializations, branches, and communications between its parallel processes/threads; (2)~The computational effort \ali{associated with each scheduled} task.

For simple applications with one or two large loops or parallel blocks of tasks that dominate its performance, inspecting \ali{the} application code \ali{is} sufficient to understand the \ali{program} flow.
If this is insufficient, tracing \ali{the} application execution can reveal the main computation and communication blocks in \ali{the} application.
In addition, the \smpi{} simulation produces a special type of text-based execution trace called time independent trace (TiT)~\cite{TiT}. 
The TiT contains a trace of the application execution as a series of computation and communication events, with their \ali{corresponding} amounts specified in \ali{terms of} \mbox{floating-pointing} operations~(FLOP) and bytes, respectively. 
\ali{Therefore,} the TiT can be used to understand the application flow and to represent the application in simulations.

To obtain the amount of work per task,  time measurement of task execution time or the FLOP count \ali{can be} used.
\ali{The} measurement of \ali{short} task execution times \ali{can be a source of measurement inaccuracies} as \ali{such measurements} are \ali{affected by the} measurement overhead which is known as the probing effect. 
In addition, \ali{the} execution time per task \fmc{is not guaranteed to} be constant between different executions of the same application.
Instead of time measurements, \ali{the} FLOP count per task can be measured using \ali{hardware counters, such as those exposed via the use of} PAPI~\cite{papi}. 
The FLOP count obtained with PAPI is used to represent the amount of work in each task in \ahmed{the} simulation. 
\ali{The} FLOP count per task is found to be a more accurate measurement  to represent computational effort per task than time measurements \ali{as well as resulting in} constant \ali{values} across different application executions~\cite{Mohammed:2018a}.
However, feeding the simulator the exact FLOP count per task might result in misrepresenting the dynamic behavior in native executions of tasks where their execution time varies \ali{among the different execution instances}.
To address this, a probability distribution \ali{is} fitted to \ali{the} measured tasks FLOP counts.
The simulator \ali{then} draws samples from this distribution \ali{to} represent the task FLOP counts during simulation as shown in the upper part of \figurename{~\ref{fig:sim_approach}}. 

\subsection{Implementing scheduling techniques for native and simulative experiments}	
\label{subsec:method_dls}
%DLS is an effective scheduling approach employed in computationally-intensive scientific applications. 
%The goal of using DLS is to optimize the performance of scientific application in the presence of load imbalance caused by problem, algorithmic, and systemic characteristics.
A number of dynamic loop \mbox{self-scheduling}~(DLS) techniques have been proposed between the late 1980s and early 2000s, and efficiently used in scientific applications~\cite{Banicescu_ETNA}.
Dynamic nonadaptive techniques have \ali{previously} been verified~\cite{HPCS} by reproduction of the original experiments that introduced them~\cite{FAC} using the experimental verification approach illustrated \ali{by step 1} in \figurename{~\ref{fig:comp_approach}}.
In this work, the range of studied DLS techniques is extended \ali{with four} adaptive DLS techniques \ali{in addition to} the nonadaptive ones.
To ensure that the implementation of \ali{the} adaptive techniques adheres to their specification, the \dlbTool{}~\cite{carino2007tool}, \ali{a dynamic load balancing library} \ali{developed} by the authors of the adaptive techniques, is used in this work.
To minimize the differences between native and simulative executions, the \dlbTool{} load balancing library, is used to schedule \ali{the application tasks} in native and simulative executions.
Connecting \ali{the} \dlbTool{} to \ali{the} simulation \ali{framework} required minimal effort \ali{as detailed} below in Section~\ref{subsec:realsim}.
\subsection{Representing native computing systems in simulation}
\label{subsec:method_system}
Representing HPC systems in simulation involves representing different system components that contribute to the application performance in simulation.
As previously investigated~\cite{Mohammed:2018a}, the application and computing system representation \ali{cannot be seen as completely decoupled activities}, i.e., representing a computing system must take into account the application characteristics as current simulators cannot simulate precisely all the complex characteristics of HPC systems to create a general, \mbox{application-independent} system representation.
For the simulation of the performance of \mbox{computationally-intensive} applications with different DLS, two main components of systems need to be represented: (1)~The PEs, their number, their computational speed; (2)~The \ali{interconnection} network between \ali{the} PEs, \ali{the} network bandwidth, \ali{the} network latency, \ali{and the} topology.

The PEs representation in simulation, \ali{needs to} reflect the \ali{native configuration in terms} of number of compute nodes and number of PEs per node. 	
Communication links connect different PEs (cores and nodes) \ali{needs to} reflect the \ali{native} network topology, \ali{bandwidth and latency}.
Nominal values for \ali{the} PE computing speeds, \ali{the} network bandwidth, and \ali{the} network latency are added in the simulated HPC representation to obtain an initial representation.
The second step is to fine tune this initial representation to reflect the ``real'' HPC performance in executing a certain application. 
To this end, core speeds are estimated to obtain more accurate simulation results due to the fact that applications do not execute at the theoretical peak performance. 
The core speed is calculated by measuring the loop execution time in a sequential run to avoid any parallelization or communication overhead. 
The sum of the total number of FLOP in all iterations is divided by the measured loop execution time to estimate the core processing speed. 
This core speed is used in the simulated HPC representation to reflect the native core speed in processing the application tasks~\cite{Mohammed:2018a}.
\alir{The above procedure is applicable for homogeneous and heterogeneous systems, where core speed estimation needs to be performed for each core type}~\cite{Mohammed:2018c}.
Similarly, a simple network benchmarking, such as \ali{a} ping-pong test was used to estimate the real network links communication bandwidth and latency and insert these values in the simulation.
Section~\ref{subsec:realsim} \ali{offers details about} the actual steps \ali{required for the calibration procedure described} above.
%\discuss{Ahmed ...system variability, how to measure, its effect, how to represent in the simulation -- do not mention implementation/simgrid related things here  ...keep the methodology abstracted from simgrid details....implementation details should be in subsection~\ref{subsec:realsim}}

Quantifying system variability is essential for achieving realistic simulations of parallel applications.
However, it involves significant challenges \ali{due to} the variety of the factors that cause the variability, e.g.,  system failures, operating system kernel interrupts, memory and network contentions~\cite{skinner2005understanding}.
The present work models the effect of the system variability on application performance  by exploiting a backlog of application execution times~\cite{bertot2018improving}.
Two factors called \textit{maximum \mbox{perturbation level}},~$PL_{max}$, and \textit{minimum \mbox{perturbation level},}~$PL_{min}$, are used to determine the upper and the lower bounds of a uniform distribution, $U$, used to estimate the \textit{\mbox{perturbation level}},~$PL$, \ali{induced by the system}. 
These factors \ali{are} calculated as in Equations~\ref{perturbation-max} and~\ref{perturbation-min}, where $E_i$ \ali{denotes} the application execution time at the $i^{th}$ \ali{execution instance} and $\bar{E}$ is the average application execution time of $n$ \ali{execution instances}.

\begin{equation}
PL_{max} = \max_{i}  \left( \cfrac{|E_i - \bar{E}| }{\bar{E}} \right) 
\label{perturbation-max}
\end{equation}

\begin{equation}
PL_{min} = \min_{i}  \left( \cfrac{|E_i - \bar{E}| }{\bar{E}} \right)
\label{perturbation-min}
\end{equation}

The estimated $PL$ is calculated as in Equation~\ref{perturbation} and is used to disturb the processor availability \ali{during} simulation, i.e.: % the performance variability \ali{observed during} a native execution of a parallel application \ali{is, therefore, injected} in the simulation.

 \begin{equation}
 PL=  U\left[PL_{min}, PL_{max} \right] 
 \label{perturbation}
 \end{equation}
 \ahmed{whenever a chunk is scheduled on a certain processor, a sample $PL$ from the uniform distribution~$U$ is drawn. 
 The value is then used to determine the speed of the processor by multiplying the original speed with ($1-P$).} 

\subsection{\fmc{Steps for Realistic Simulations}}
\alir{To achieve realistic performance simulation, three factors that affect application performance need to be \fmc{well} represented. 
In this section, we summarize the steps of the proposed realistic simulation approach and different methods to represent each factor.

\begin{enumerate}
	\item[\textbf{Step 1}] Application characteristics
	\begin{enumerate}
		\item Program flow
		\begin{itemize}
			\item \fmc{Study the application source code or}
			\item \fmc{Trace the execution of the application} 
		\end{itemize}
		\item Computational effort per task
		\begin{itemize}
			\item \fmc{Collect time measurements} for tasks of large granularity,
			\item \fmc{Measure the} FLOP count per task (large- or fine-grain tasks), \fmc{or}
			\item \fmc{Use a FLOP probability} distribution to capture variability in native executions
		\end{itemize}
	\end{enumerate}
	\item[\textbf{Step 2}] Task scheduling
	\begin{enumerate}
		\item Implement and verify scheduling techniques in the simulator \fmc{or}
		\item Use the native library to schedule tasks in simulation, similar to the native tasks
	\end{enumerate}
	\item [\textbf{Step 3}] Computing system characteristics
	\begin{enumerate}
		\item PEs representation
		\begin{itemize}
			\item \fmc{Represent each PE} in simulation to have full control on its behavior in simulation
			\item Estimate core speed by dividing application execution time by the FLOP count of the application \fmc{and}
			\item Cores that represent \fmc{a single} node should be connected to each other by simulated links that represent memory bandwidth and latency
			\end{itemize}
		\item Interconnection network
		\begin{itemize}
			\item \fmc{Represent the network topology of} the simulated system
			\item \fmc{Use a network model in simulation that captures} the characteristics of the native interconnection fabrics (e.g., InfiniBand)
			\item Use nominal network link bandwidth and latency \fmc{values}
			\item Fine tune this representation by running network benchmarks and adjust bandwidth, latency, and other delays for large and small messages
		\end{itemize}
		\item System variability
		\begin{itemize}
			\item Model variations in applications execution time as \fmc{independent and uniformly distributed random variables}
			\item Draw samples from the \fmc{uniform} distribution to change the availability of system components during simulation
		\end{itemize}
	\end{enumerate}
\end{enumerate}

}
%\begin{equation}
%\label{perturbation-min}
%
%\end{equation}

% !TEX root =  fgcs_19.tex
\section{Experimental Evaluation and Results}  
\label{sec:ex}
\ali{To evaluate the usefulness and effectiveness of the proposed approach, an important number of native and simulative experiments is \fmc{performed}.
These experiments have been designed as a factorial set of experiments which is described below and summarized in Table~\ref{tbl:exp}.}
%The factorial design of experiments is explained in this section and summarized in Table~\ref{tbl:exp}.
In addition, details of creating the performance simulation using \simgrid{} and its interfaces and how the \fk{approach} proposed in Section~\ref{sec:methodology} is applied to realistically simulate the performance of two scientific \mbox{computationally-intensive} applications \fmc{are} \ali{also provided}.
\ali{Subsequently}, \ali{the} native and simulative performance results are compared using the second and the third step of \ali{the} comparison approach illustrated in~\figurename{~\ref{fig:comp_approach}} and the results are discussed.

\subsection{Design of native and simulative experiments}
\label{subsec:designofexp}
\begin{table}[!h]
	\caption{Details used in the design of factorial experiments}
	\label{tbl:exp}
	\centering
	\begin{adjustbox}{max width=\textwidth}
		\begin{tabular}{@{}lll@{}}
			\toprule
			\textbf{Factors}             & \textbf{Values}                                                                                                                  & \textbf{Properties}                                                                                                                                                               \\ \midrule
			\multirow{2}{*}{\textbf{Applications}}                                                                                                                                      & PSIA                                                                                                                                                                                                                    & \begin{tabular}[c]{@{}l@{}} $N = 400000$ tasks \\ Low variability among tasks\end{tabular}                                                                                                                                                                                                                    \\ \cmidrule(l){2-3} 
			& Mandelbrot                                                                                                                                                                                                              & \begin{tabular}[c]{@{}l@{}} $N = 262144$ tasks\\ High variability among tasks \end{tabular}                                                                                                                                                                                                                  \\ \midrule
			\multirow{3}{*}{\begin{tabular}[c]{@{}l@{}}\textbf{Self-scheduling} \\ \textbf{techniques}\end{tabular}}                                                                                        & STATIC                                                                                                                                                                                                                  & Static                                                                                                                                                                                                                                                                                                   \\ \cmidrule(l){2-3} 
			& \begin{tabular}[c]{@{}l@{}}mFSC, GSS, FAC\end{tabular}                                                                                                                                             & \begin{tabular}[c]{@{}l@{}}Dynamic nonadaptive\\ \end{tabular}                                                                                                                                                                                                \\ \cmidrule(l){2-3} 
			& AWF-B, -C, -D, -E                                                                                                                                                                                                  & \begin{tabular}[c]{@{}l@{}}Dynamic adaptive\\ \end{tabular}                                                                                                                                                                                                   \\ \midrule
			\textbf{\begin{tabular}[c]{@{}l@{}}Computing\\ system\end{tabular}}    & miniHPC                                                                                                                          & \begin{tabular}[c]{@{}l@{}}16 Dual socket Intel $E5-2640 v4$ nodes\\ 10 cores per socket\\$64$~GB \ahmed{DDRAM per node} \\Nonblocking fat-tree topology\\ \ahmed{Fabric: Intel OmniPath - 100~Gbps}        \end{tabular} \\ \midrule
			\multirow{2}{*}{\textbf{\begin{tabular}[c]{@{}l@{}} \\ Experimentation \end{tabular}}}	& Native &\begin{tabular}[c]{@{}l@{}}\ahmed{$P = 16, 32, 64, 128, 256$ PEs} \\\ahmed{using $1, 2, 4, 8, 16$ miniHPC nodes, 16 PE per node} 
			\end{tabular}\\ \cmidrule(l){2-3} 
			& \begin{tabular}[c]{@{}l@{}}  Simulative \end{tabular}                                                & \begin{tabular}[c]{@{}l@{}}  \ahmed{$P = 16, 32, 64, 128, 256$} simulated PEs\\
				\ahmed{using $1, 2, 4, 8, 16$ simulated miniHPC nodes}, 16 PE per node \\ (1)~Using \emph{FLOP file} with \smpi{}+\msg{} \\ (2)~Using \emph{FLOP distribution} with \smpi{}+\msg{} \end{tabular}    \\ \bottomrule
		\end{tabular}
		
	\end{adjustbox}
\end{table}

%\begin{itemize}
%\item applications - PSIA and Mandelbrot 
%\item Scheduling - all 9 DLS available in \dlbTool{}
%\item HPC - miniHPC - Xeon and KNL 
%\end{itemize}

\begin{figure}[!h]
	\centering
	\includegraphics[clip, trim=0cm 0cm 0cm 0cm, scale=1.2]{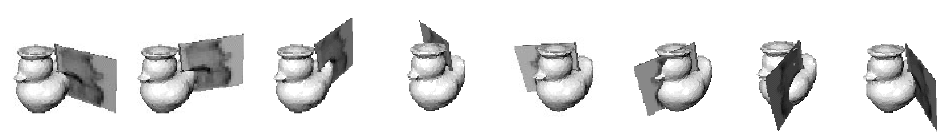}%
	\caption{Illustration of the spin-image calculation for a 3D object~(from literature~\cite{sia}). A flat sheet is rotated around each point of the 3D object to describe the object from this point view.}
	\label{fig:PSIA}
\end{figure}

\noindent\textbf{Applications.}
The first application considered in this work is \fmc{the parallel spin-image algorithm~(PSIA), a \mbox{computationally-intensive} application from computer vision}~\cite{psia}.
The \fmc{core computation of the sequential version of the algorithm (SIA)} is the generation of the 2D \mbox{spin-images}.
\figurename{~\ref{fig:PSIA}} shows the process of generation of a \mbox{spin-image} for a 3D object.
The PSIA exploits the fact that \mbox{spin-images} generations are independent of each other.
The size of a single \mbox{spin-image} is small (200~bytes) and fits in the lower level (L1) cache.
Therefore, the memory subsystem has \alir{no} impact on the application performance, \alir{as data are always available for computation at the highest speed}.
The \fmc{PSIA} pseudocode \ali{is available} online~\cite{ISPDC_online}.
The amount of computations \ali{required} to generate \ali{the} spin-images is data-dependent and \ali{is} not identical over all the \mbox{spin-images} generated from the same object.
This introduces an algorithmic source of load imbalance among the parallel processes generating the spin-images.
The performance of PSIA has previously \ali{been} enhanced by using nonadaptive DLS techniques to balance the load between the parallel processes~\cite{Eleliemy:2017b}.
Using DLS improved the performance of the PSIA by a factor of~$1.2$ and~$2$ for homogeneous and heterogeneous computing systems.

\begin{figure}[!h]
	\centering
	\includegraphics[clip, trim=0cm 0cm 0cm 0cm, scale=0.32]{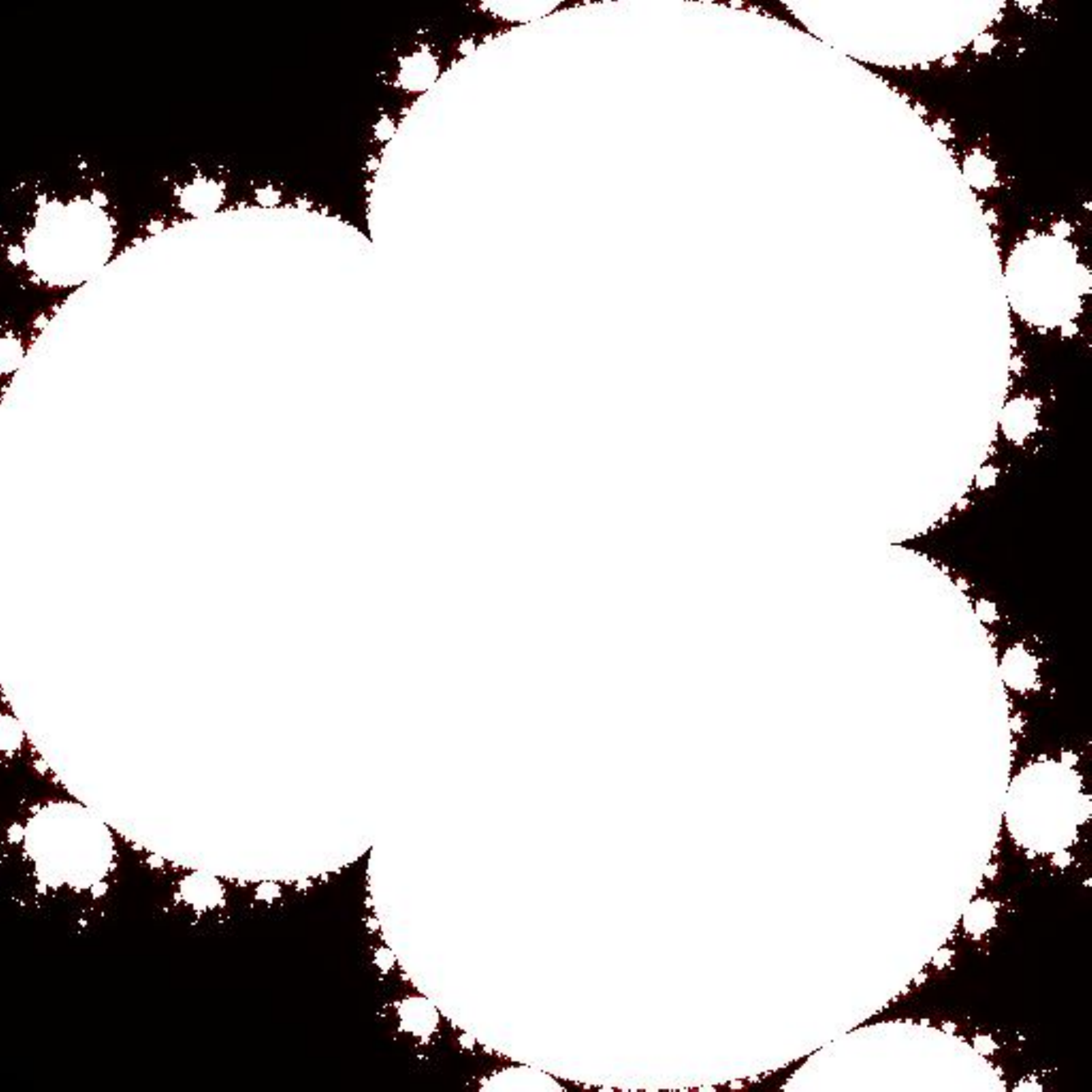}%
	\caption{Mandelbrot calculation at the seahorse valley for $z^4$. White points represent high computational load \ali{due to several iterations to reach convergence} and black points represent \ali{negligible} computations \ali{whereby saturation is reached in a few iterations}.}
	\label{fig:mandel_center}
\end{figure}

The second application of interest is the \fmc{computation of the} Mandelbrot set~\cite{mandelbrot1980fractal} and the \fmc{generation of its corresponding} image. 
The application is parallelized such that the calculation of the value at every single pixel of a 2D image is a loop iteration, that is performed in parallel. 
The application computes the function $f_c(z) = z^4 + c$ instead of $f_c(z) = z^2 + c$ to increase the number of computations per task. 
The size of the generated image is $512 \times 512$ pixels resulting in $2^{18}$ parallel loop iterations.
To increase the variability between tasks execution times, the \ali{calculation} is focused on the center image, i.e., the seahorse valley, where the computation is \mbox{intensive}.
\figurename{~\ref{fig:mandel_center}} shows the calculated image.
Mandelbrot is often used to evaluate the performance of dynamic scheduling techniques due to the high variation between its loop iterations execution times.\\

\noindent\textbf{Dynamic load balancing.}
The \dlbTool{} is an \mbox{MPI-based} dynamic load balancing library~\cite{carino2007tool}.
The \dlbTool{} has been used to balance the load of scientific applications, such as image denoising and the \ali{statistical analysis of} vector nonlinear time~\cite{carino2007tool}.
The \dlbTool{} is used for the self-scheduling of the parallel tasks of PSIA and Mandelbrot both in native and simulative executions.
The \dlbTool{} employs a master-worker execution model, where the master also \fk{acts} as a worker when it is not serving worker requests.
Workers request work from the master whenever they become \fk{idle}, \ali{i.e., the} \mbox{self-scheduling} \ali{work distribution}.
Upon receiving a work request, the master calculates a chunk size based on the used DLS technique. 
Then, the master sends the chunk size and the start index of the chunk to the requesting worker.
The above process of work requests from workers and master assigns work to requesting workers repeats until the work is finished.
The \ali{two applications of interest are scheduled using the \dlbTool{} with eight different loop scheduling techniques} ranging from static to dynamic, nonadaptive and adaptive as shown in Table~\ref{tbl:exp}. 

\noindent\textbf{Computing system.}
The miniHPC\footnote{https://hpc.dmi.unibas.ch/HPC/miniHPC.html} is a high performance computing cluster \ali{at} the Department of Mathematics and Computer Science at \ali{the} University of Basel, Switzerland, \ali{used for for research and teaching}.
%It consists of 26~multicore compute nodes and \fmc{two dedicated nodes for login and storage}.  
%The miniHPC cluster has a theoretical peak performance of $30~\mathit{TFLOP/s}$. 
%\fmc{Twenty-two nodes} \ali{consist of} two Intel Broadwell CPUs.
For the experiments in this work, $16$~dual-socket nodes are used, \fmc{where each socket holds an Intel Broadwell CPU with 10 cores}. 
%Four compute nodes \fmc{contain each a} \mbox{standalone} Intel Xeon Phi processor.
The hardware characteristics of the miniHPC \fmc{nodes} are listed in Table~\ref{tbl:exp}.
All nodes are connected via Intel \mbox{Omni-Path} interconnection fabric in a nonblocking \mbox{two-level} \mbox{fat-tree} topology.
The network bandwidth is $100~\mathit{Gbit/s}$  and the communication latency is $100~\mathit{ns}$.

\subsection{Realistic simulations of scientific applications}
\label{subsec:realsim}
\noindent\textbf{Extracting the computational effort in an application.}
To obtain the computational effort per task of the applications of interest, the FLOP count \ali{approach described in Section~\ref{subsec:method_app}} is used.
The native application code is instrumented and the number of FLOP per task is counted using the PAPI \fmc{performance API}~\cite{papi}.
The application was \ali{executed} sequentially on a single dedicated node in the FLOP counting experiment to avoid interference between cores on the hardware counters and ensure the correct count of FLOPs.
The experiment was repeated $20$ times for each application to ensure that the \ali{FLOP} count is \ali{constant} in all repetitions.
\ali{The} FLOP count can be also inferred from the application source code~\cite{HPCS} in case of simple dense linear algebra kernels.
The resulting FLOP count per task is written to a file that is read by the simulator to account for task execution times.
\ali{Whenever} inferring or counting FLOP per task is not possible, and tasks are of large granularity, \ali{the} task execution time \ali{can} be used instead of FLOP count, as the measurement overhead \ali{will not dominate the task execution time as it is the case for short tasks}.

\begin{figure}[!h]
	\begin{minipage}{\textwidth}
		%	\begin{adjustbox}{minipage=\linewidth,frame}
		\centering
	
			\begin{tabular}{@{}cc@{}}
			
				\includegraphics[clip, trim=1cm 6cm 2cm 6cm, scale=0.345]{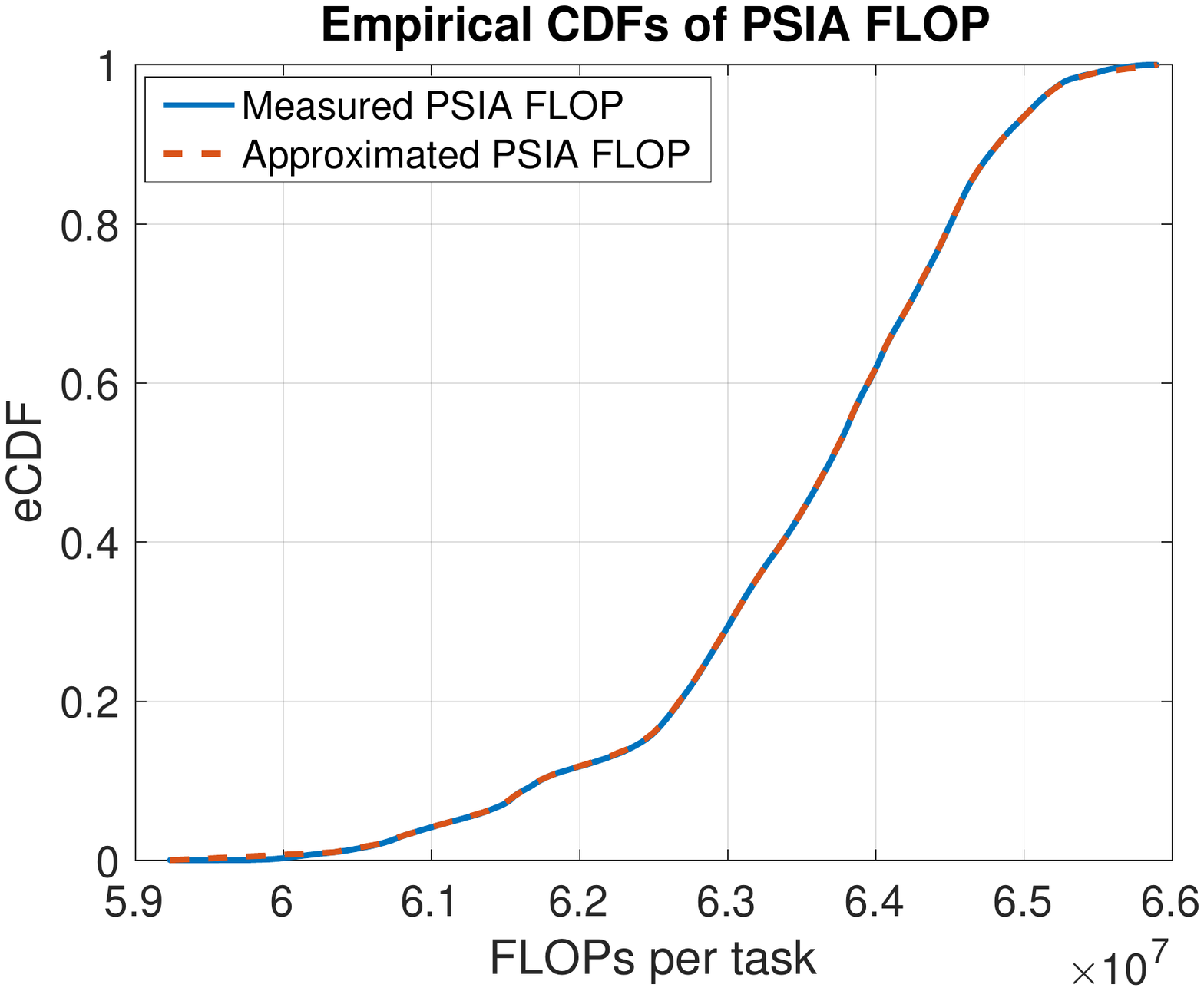}	& 	\includegraphics[clip, trim=1cm 6cm 2cm 6cm, scale=0.345]{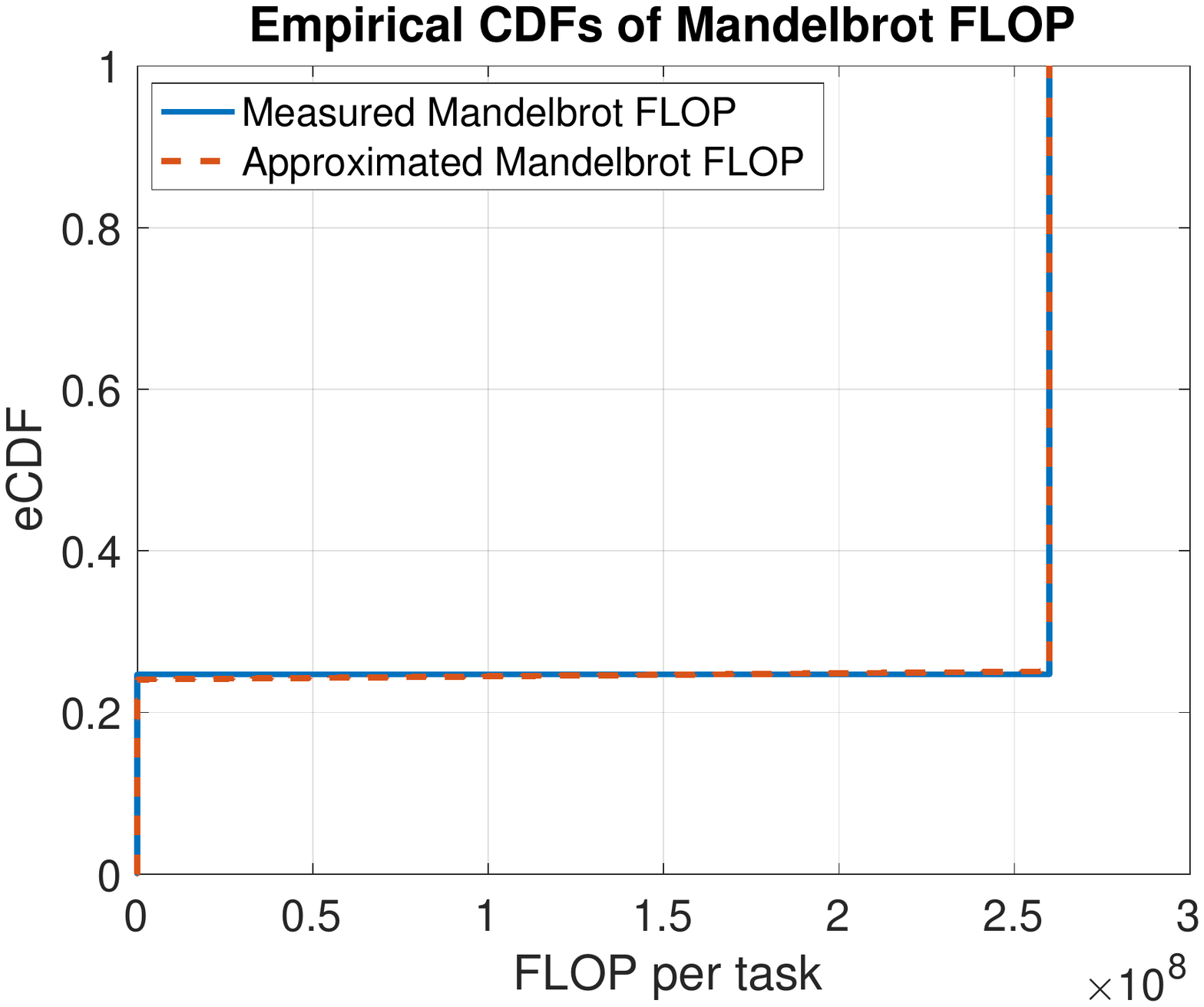} \\ 
			{\small (a) eCDF of PSIA tasks FLOP count}	& {\small (b)~eCDF of Mandelbrot tasks FLOP count}  \\ 
			\end{tabular}
	   \end{minipage}
		\caption{Empirical cumulative density function of the tasks FLOP counts of PSIA and Mandelbrot. The distribution of the measure FLOP count is shown in blue and the distribution of the FLOP counts drawn from the linear piecewise approximation of the eCDF is shown in orange. The results show that approximated distribution represents the empirical measure FLOP counts of both applications closely.  }
		\label{fig:FLOP_approx}
	%	\vspace{-0.5cm}
\end{figure}

To simulate the dynamic behavior of \ali{the} task execution times, \ali{a} probability distribution is fitted to the \ali{measured} FLOP count. 
To obtain this probability distribution, \ali{the} linear piecewise approximation of the empirical cumulative density function~(eCDF) is used~\cite{beaumont2018influence}. 
The eCDF \ali{values} are split over the \mbox{y-axis} into $100$~linear segments~(pieces). 
To draw a sample from this distribution, a segment is randomly selected, and a value is randomly selected along this linear segment.
\figurename{~\ref{fig:FLOP_approx}} shows the results of approximating the measured FLOP counts of tasks \ali{both} from PSIA and Mandelbrot using linear piecewise approximation of the eCDF using MATLAB\footnote{https://www.mathworks.com/products/matlab.html}.
To ensure that the simulator draws samples from \ali{the} approximated distribution with a \ali{fast, long period, and low serial correlation} random engine, the random number generator of the GNU Scientific Library\footnote{https://www.gnu.org/software/gsl/doc/html/index.html}~(GSL) is used in the simulator to generate good uniformly distributed random numbers to select \ali{among} the $100$ linear segments and a value from the segment with low overhead during simulation.
 
%%Two simulation approaches are experimented in this work using different interfaces of SimGrid.
%\noindent\textbf{MSG simulation.} \discuss{Franziska}\\

\noindent\textbf{The SMPI+MSG simulation approach.}\\
A novel simulation approach is employed in this work. 
Two interfaces of the SimGrid toolkit \ali{are} leveraged to realistically simulate the application performance with minimal effort.
\ali{Algorithm 1 shows the changes needed in the native application code to transform it into the simulative application code using \smpi{}+\msg{} using the \fk{approach} illustrated in \figurename{~\ref{fig:sim_approach}}.}
Lines in \ali{mint font color in Algorithm~\ref{algo:smpi_msg} show additions to simulate the application, lines in grey font color show the lines that need to be uncommented to revert to the native application code, and black lines denote unchanged code.}
\begin{algorithm}[!h]
	\small
	\caption{\small Native code transformation into SMPI+MSG simulative code}
	\label{algo:smpi_msg}
	\#include $<$mpi.h$>$\\
	{\#include ``DLB\_tool.h''}\\
	{\color{UnibasMint} \#include ``msg.h'' /* simulative only*/}\\
	
	MPI\_Init(\&argc, \&argv);\\
	MPI\_Comm\_size(MPI\_COMM\_WORLD, \&$P$);\\ 
	MPI\_Comm\_rank(MPI\_COMM\_WORLD, \&myid);\\
	
	/* Initialization */ \\
	\dots\\
	{\color{UnibasGrey} /* results\_data = malloc($N$); native only*/} \\
	{\color{UnibasMint} tasks = create\_MSG\_tasks($N$); /* simulative only */} \\
	{DLS\_setup(MPI\_COMM\_WORLD, DLS\_info);} \\
	{ DLS\_startLoop (DLS\_info, $N$, DLS\_method);}\\
	
	t1 = MPI\_Wtime();\\
	\While{ { Not DLS\_terminated}}
	{
		DLS\_startChunk(DLS\_info, start, size);\\
		{\color{black} 
			/* Main application loop */\\
			{\color{UnibasGrey}\textbf{/* Compute\_tasks(start, size, data); native only */}}\\
			{\color{UnibasMint} \textbf{Execute\_MSG\_tasks(start, size);} /* simulative only */}\\
		} %end black
		{ DLS\_endChunk(DLS\_info);}\\
	} %end while
	
	DLS\_endLoop(DLS\_info);\\
	t2 = MPI\_Wtime();\\
	print("Parallel execution time: \%lf \textbackslash{}n", t2 - t1);\\
	{\color{UnibasGrey} /* Output or save results removed from simulation- native only */}\\
	\dots\\
	{ MPI\_Finalize();} \\  
\end{algorithm}	

The \smpi{} interface is used to execute the native application code.
To speedup the \smpi{} simulation, the computational tasks in the application are replaced with \msg{} tasks.
The amount of work per \msg{} task is either read from a file or drawn from a probability distribution according to the experimented simulation type.
Memory allocations of results and data in the native code are removed or commented in the simulation as they are not needed.
This allows to reduce the memory footprint of the simulation and the simulation of a large number of ranks on a single compute node.  
No modifications are needed for the \dlbTool{} in this approach.
The scheduling overhead of different techniques is accounted for by the \smpi{}, whereas the tasks execution time is accounted for in simulation by the \msg{}.
The proposed approach results in a fast and accurate simulation of the application with minimal modifications to the native application source code.
Hundreds \ali{to} thousands of MPI ranks can be simulated using a single core on a single compute node. \\

\noindent\textbf{Computing system representation.}\\
To represent the miniHPC in SimGrid, the system characteristics need to be entered in a specially formatted XML file denoted as \texttt{platform file}.
Each core of a compute node of miniHPC is represent as a host in the \texttt{platform file}.
Hosts that represent the cores of the same node are connected with links with high bandwidth and low latency to represent communication of cores of the same node through the memory.
The bandwidth and the latency of these links are used as $500~\mathit{Mbit/s}$ and $15~\mathit{us}$, respectively to represent the memory access bandwidth and latency.
Every $16$ host represent a node of miniHPC.
Another set of links are used to connect the hosts to represent network communication in a \mbox{two-level} \mbox{fat-tree} topology.
The properties of these links represents the properties of the Intel Omni-Path interconnect used in miniHPC and their bandwidth and latency are set to $100~\mathit{Gbit/s}$ and $100~\mathit{ns}$, respectively.

To reflect the fact that network communications are nonblocking in the native miniHPC system, the \texttt{FATPIPE} is used to tell SimGrid that the communications on these links are nonblocking and is not shared, i.e., each host has all network bandwidth and shortest latency available all the time even in the case of all hosts are communicating at the same time.
For the links that represent the memory communication, their sharing property is set to \texttt{SHARED} to represent possible delays that can occur if multiple cores are trying to access the memory at the same time.

To estimate the core speed, each application is \ali{executed} sequentially on a single core to estimate the total execution time and avoid any scheduling or parallelization overhead in this measurement.
The core speed is calculated as the total number of FLOP in all tasks of the application divided by the total application sequential execution time.
Using the above approach, the core speed is found to be $0.95~\mathit{GFLOP/s}$ and $1.85~\mathit{GFLOP/s}$ for the execution of PSIA and Mandelbrot, respectively.
This requires the creation of two \texttt{platform files} to represent the miniHPC in the execution of PSIA and Mandelbrot. 
This illustrates the strong coupling between application and system representation in simulation as discussed in Section~\ref{subsec:method_system}.

\alir{
\mbox{SimGrid} uses a \mbox{flow-level} \fmc{network modeling} approach that realistically approximates the behavior of TCP and InfiniBand~(IB) networks specifically tailored for HPC settings\ahmed{.}
\ahmed{This approach} accurately models contention in \ahmed{such} networks~\cite{vienne2010prediction} \ahmed{and} accurately captures the network behavior for messages larger than $100KB$ on highly contended networks~\cite{velho2009accuracy}\ahmed{.}
The \mbox{SimGrid} \ahmed{network} model can further \ahmed{be} configured to precisely capture characteristics, such as \fmc{the} slow start of MPI messages, \ahmed{\mbox{cross-traffic}}, and asynchronous send~\footnote{https://simgrid.org/doc/latest/Configuring\_SimGrid.html\#options-model-network} \ahmed{calls}.}
To fine tune the network \fmc{representation} in the simulation to the native miniHPC system, the \simgrid{}-based calibration procedure~\cite{simgrid_calib} is used to calibrate \alir{the network model parameters in} the representation of both platforms to better adjust the network bandwidth and latency in both \texttt{platform files}. 

Using the approach introduced in earlier work~\cite{Ali_SC:17}, the representation of the computing system can be verified in a separation of the application representation by using the \smpi{} interface.
The \smpi{} interface simulates the execution of native MPI codes on a simulated computing \texttt{platform file}. Both the native and simulative executions using \smpi{} share the application’s native code. 
The difference between the native execution and the simulative \smpi{}-based execution is the computing system representation component. 
The representation of the computing system can be verified by comparing the native and \smpi{} simulative performance results.

To quantify the effect of system variability, both applications, PSIA and Mandelbrot, were executed 20 times using STATIC on \ahmed{256 PEs.}
For PSIA, $\bar{E}$, $PL_{max}$, and $PL_{min}$ were $111.5792$ seconds, $0.1539$, and  $0.0113$, respectively. 
For Mandelbrot, $\bar{E}$, $PL_{max}$, and $PL_{min}$ were $139.9814$ seconds, $0.0088$, and $0.0009$, respectively. 
\ahmed{These} results indicate a low system variability \ahmed{in} miniHPC \ahmed{during the execution of} both applications\ahmed{.}
\ahmed{This variation is not considered in the simulative experiments.}

%\begin{itemize}
%	\item how we represented the xeon partition of miniHPC
%	\item estimated core speeds and network
%	\item platform calibration
%	\item \discuss{Ahmed: measured system variability on miniHPC xeon and KNL - represent/ignore in simulation}
%\end{itemize}
\subsection{Experimental results}
\label{subsec:results}
\begin{figure}[]
	\begin{minipage}{\textwidth}
		%	\begin{adjustbox}{minipage=\linewidth,frame}
		\centering
		
		\begin{tabular}{@{}cc@{}}
			
			\includegraphics[clip, trim=0cm 0cm 0cm 0cm, scale=0.315]{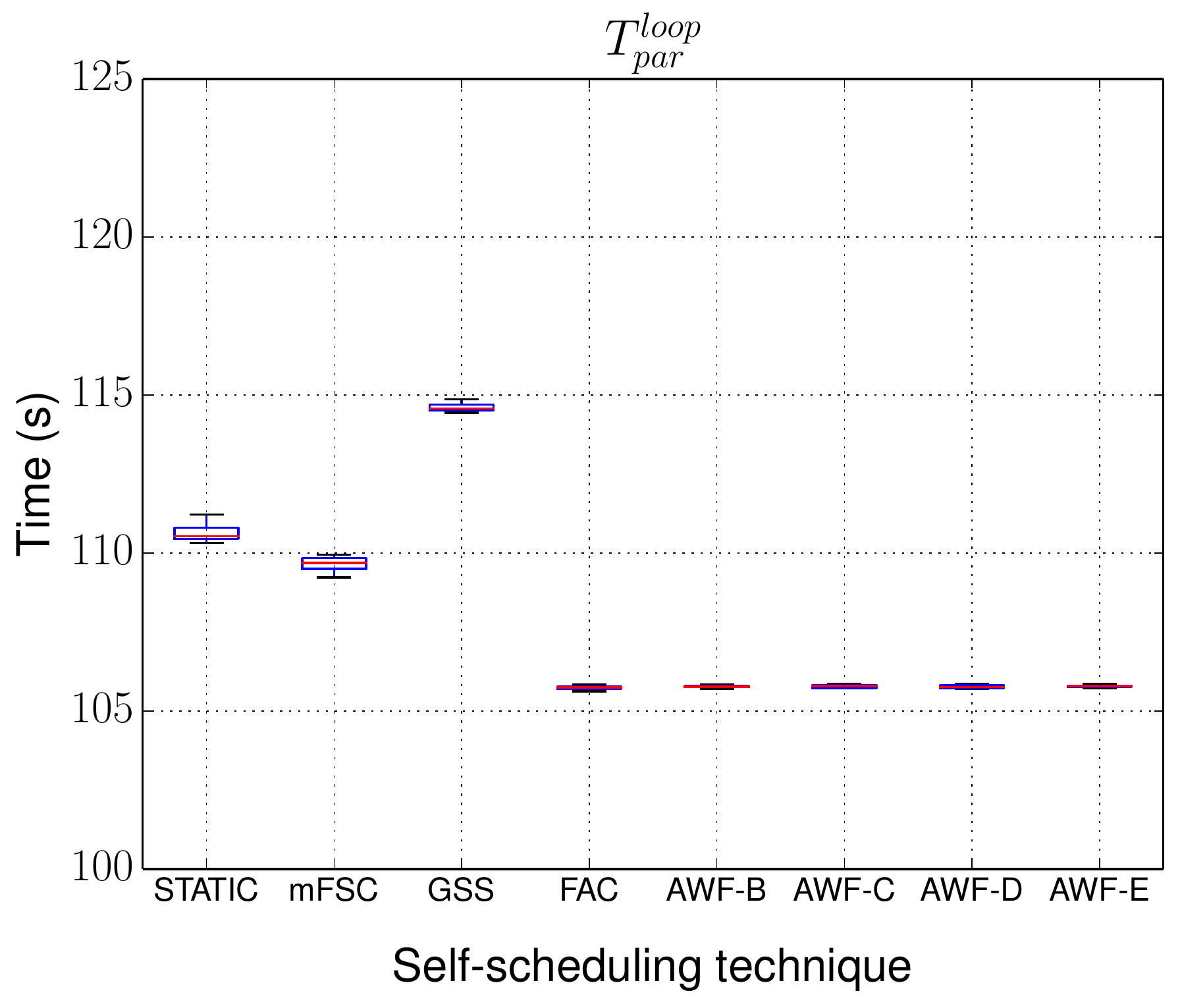}	& 	\includegraphics[clip, trim=0cm 0cm 0cm 0cm, scale=0.315]{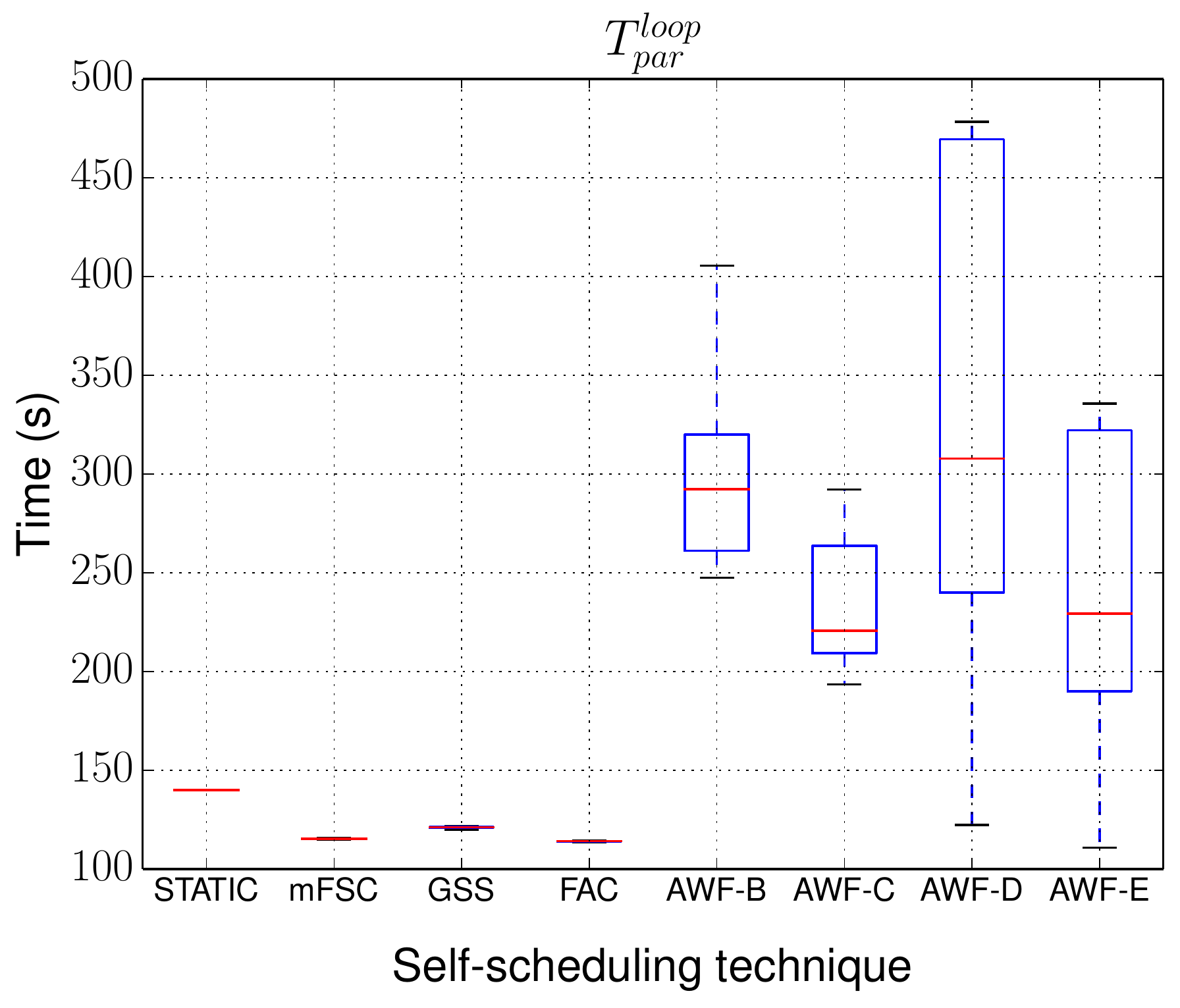} \\ 
			{\small (a) PSIA native performance}	& (b) {\small Mandelbrot native performance } \\

			\includegraphics[clip, trim=0cm 0cm 0cm 0cm, scale=0.315]{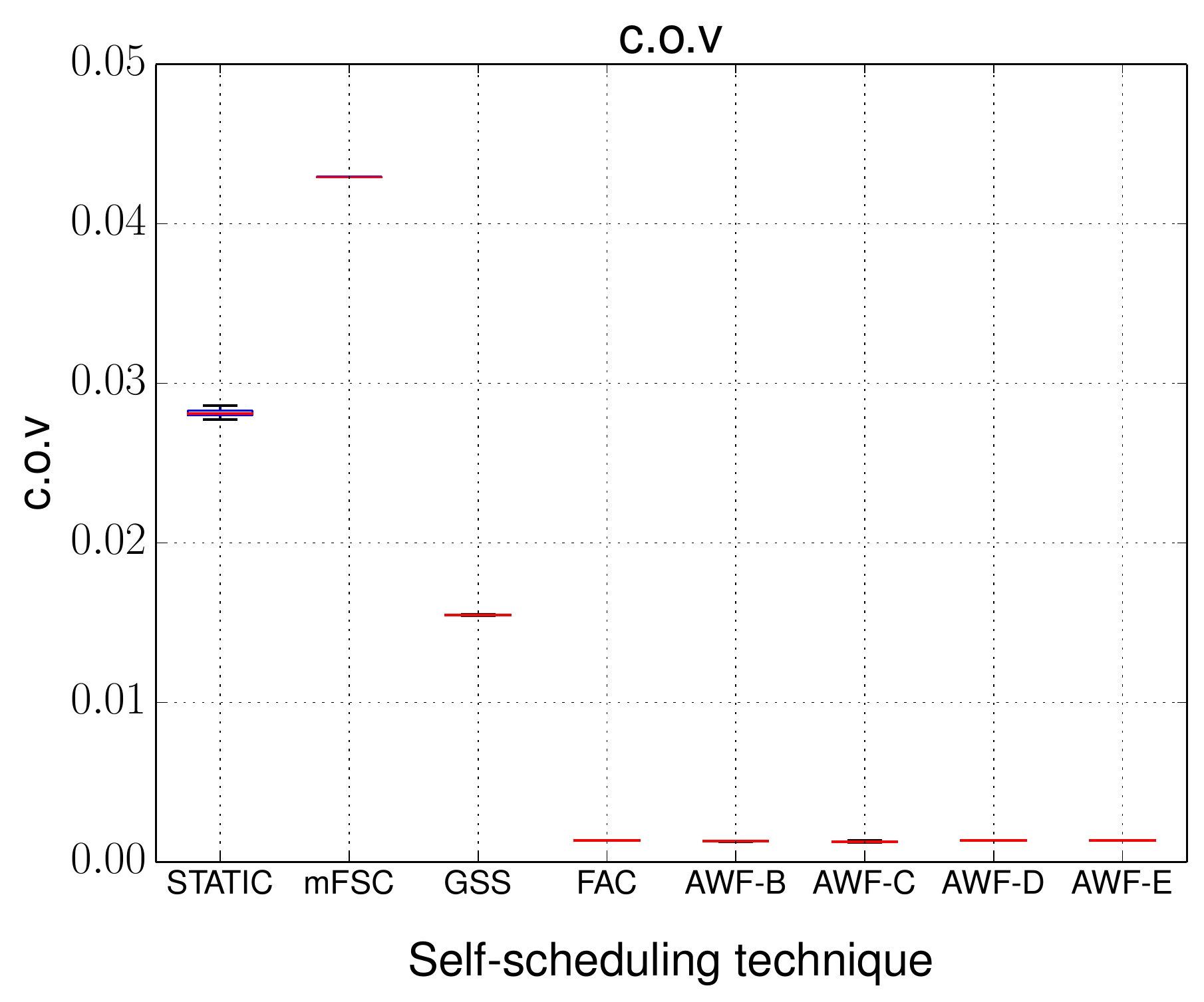}	& 	\includegraphics[clip, trim=0cm 0cm 0cm 0cm, scale=0.315]{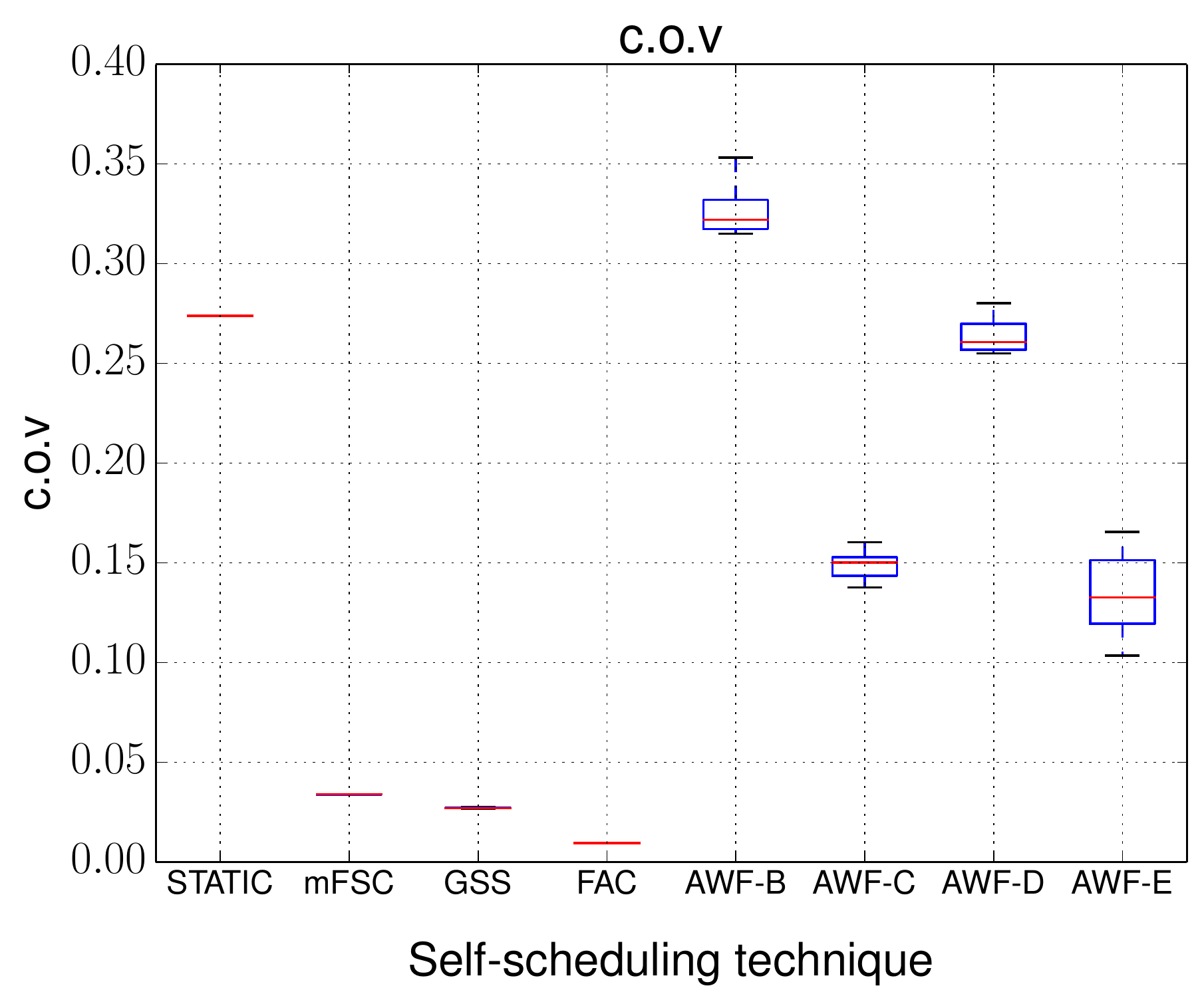} \\ 
			{\small(c) Load imbalance in PSIA (c.o.v.)}	& {\small(d)~Load imbalance in Mandelbrot (c.o.v.) } \\

			\includegraphics[clip, trim=0cm 0cm 0cm 0cm, scale=0.315]{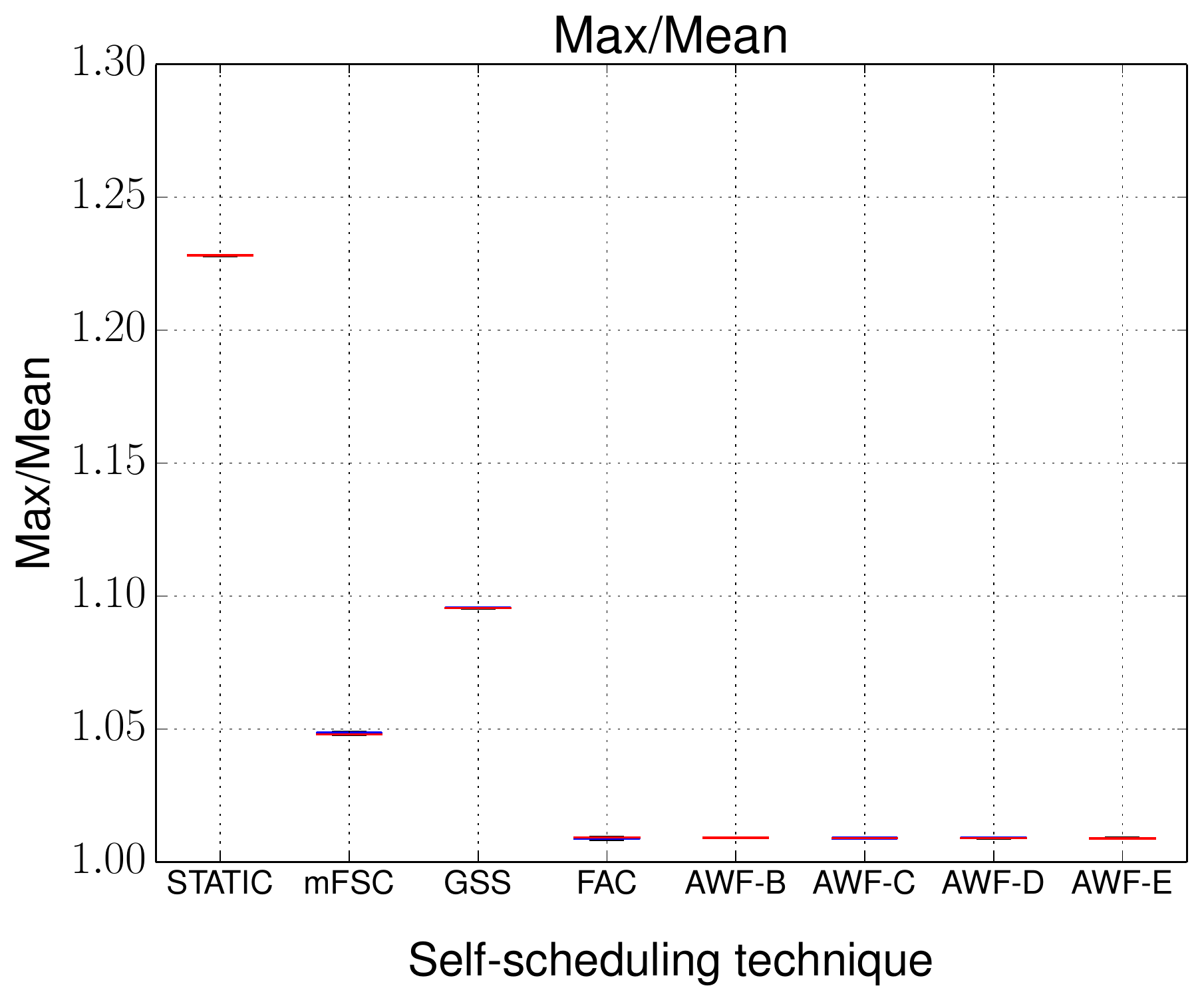}	& 	\includegraphics[clip, trim=0cm 0cm 0cm 0cm, scale=0.315]{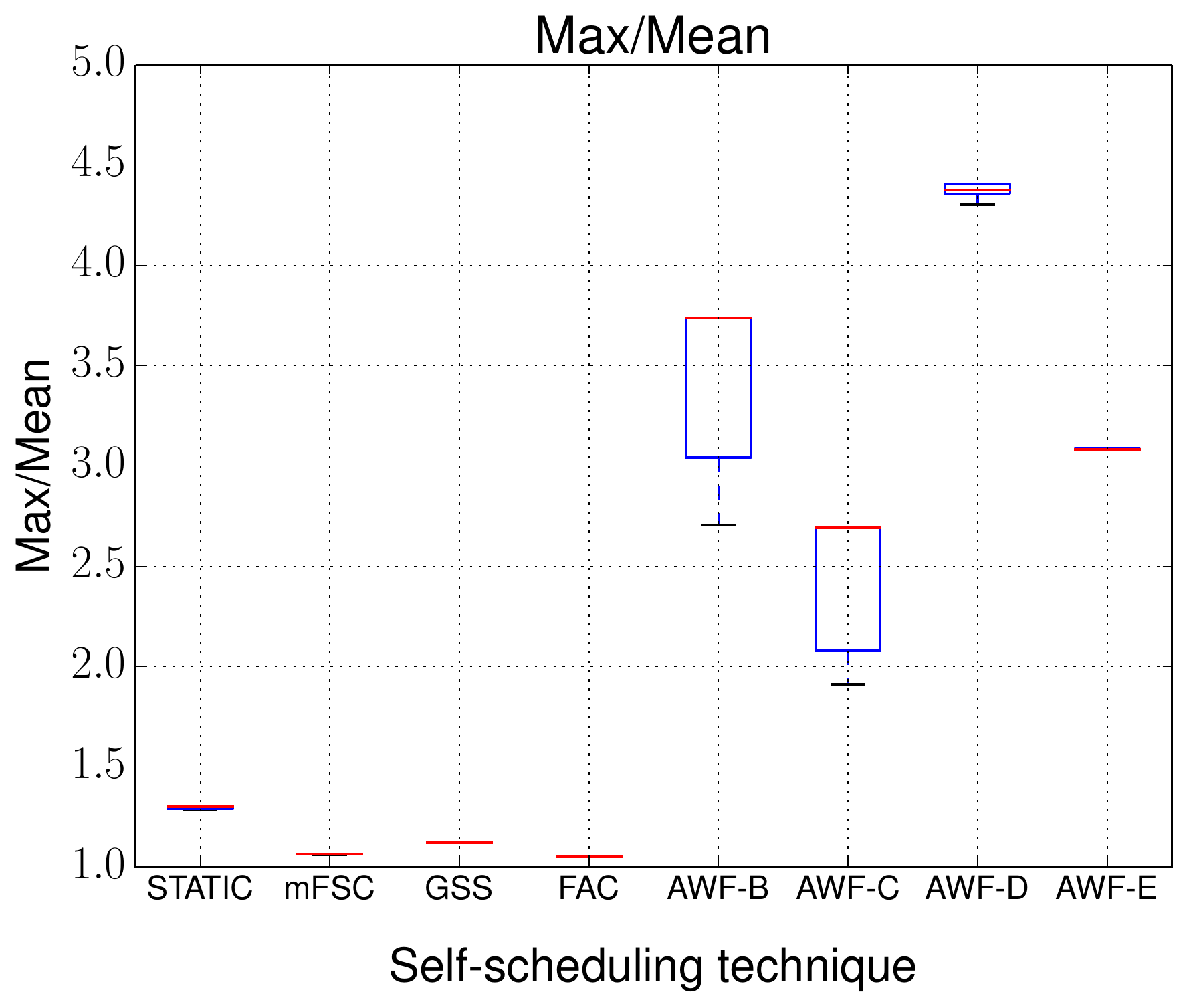} \\ 
			{\small (e) Load imbalance in PSIA (max/mean)} 	& {\small (f) Load imbalance in Mandelbrot (max/mean) }  \\ 
		\end{tabular}
	\end{minipage}
    \caption{\textbf{Native} performance of PSIA and Mandelbrot applications. STATIC degrades applications performance due to high load imbalance. Applications performance improves with FAC. Adaptive techniques improve the performance of PSIA; however, they degrade Mandelbrot performance and do not adapt correctly.}
	\label{fig:native}
	%	\vspace{-0.5cm}
\end{figure}

\figurename{~\ref{fig:native}} shows the native performance of both PSIA and Mandelbrot with \ali{eight} static \ali{and} dynamic (nonadaptive and adaptive) \mbox{self-scheduling} techniques.
To measure application performance, the parallel loop execution time $T^{loop}_{par}$ for both applications is reported.

\alir{
Each \ahmed{native} experiment is \ahmed{submitted for execution} as a single job \ahmed{to} the Slurm~\cite{yoo2003slurm} \ahmed{batch} scheduler on dedicated miniHPC nodes.
\ahmed{Slurm exclusively allocates nodes to each job.}
The nonblocking \mbox{fat-tree} network topology of miniHPC guarantees that nodes use the full bandwidth of \ahmed{the links,} even if other applications are running on other nodes in the cluster.
\ahmed{The a}pplication codes are compiled with \ahmed{the} Intel compiler v. 17.0.1 \ahmed{without any compiler optimizations} to prevent compiler from changing the applications.
\ahmed{Such changes in application behavior would have undesired consequences in the fidelity of the application representation in simulation.}
miniHPC runs the CentOS Linux version~7 operation system.
}

Each \ahmed{native} experiment is repeated $20$ times to obtain performance results with high confidence.
The boxes represent the first and third quartiles, the red line represents the median of the $20$ measurements, and the whiskers represent $1.5 \times$ the standard deviation of the results.

Two metrics are used to measure the load imbalance in both applications: (1)~the \emph{coefficient of variation} (c.o.v.) of the processes finishing times~\cite{FAC} and (2)~max/mean of the processes finishing times.

The c.o.v. is calculated as the standard deviation of processes finishing times divided by their mean and \fmc{indicates load imbalance as the} variation in general between the processes finishing time.
A high \ali{c.o.v.} value represents high load imbalance and a low value (near zero) represents a \fmc{nearly perfectly} balanced load execution.

The max/mean is calculated as the maximum of processes finishing times divided by their mean.
Max/mean indicates how long the processes of an application had to wait for the slowest process due to load imbalance. 
A \ali{max/mean} value of $1$ represents a balanced load execution (lower bound), and a higher value indicates that execution time is prolonged due to a process that lags all the other processes at the end.

\fmc{\ali{When} all processes\fk{,} except for one\fk{,} have similar finishing times, the c.o.v. is very low and hides the fact that the slowest process lags behind in execution, while the finishing time of this process is visible as a large value \ali{in} max/mean metric.}

Inspecting the native applications results in \figurename{~\ref{fig:native}}, one observes that STATIC degraded the performance of both PSIA and Mandelbrot due to load imbalance.
The high value of c.o.v and max/mean in both applications indicate the load imbalance with STATIC as shown in \fk{subfigures} \fk{(c) and (d)}.
Although the value of c.o.v for GSS is lower than that of mFSC for PSIA, one can see that the performance of GSS is worse than mFSC. 
\figurename{~\ref{fig:native}} \fk{(e)} shows, however, that the value of max/mean for GSS is higher than that of mFSC, which explains the large execution time in \fk{subfigure} (a). 
This is an example where the c.o.v. alone hides the load imbalance resulting from a single process lagging the application execution as explained above.
FAC technique improves the performance of both applications and result in the lowest execution time and also load imbalance metrics.

\ali{The} adaptive \fmc{DLS} techniques improve the performance of PSIA and result in low load imbalance metrics as well.
However, for the Mandelbrot due to the high variability of its tasks execution times and short execution times, the adaptive techniques did not have enough time to estimate PE relative weights correctly and resulted in high execution time and high load imbalance metric values with high variability also.

\begin{figure}[]
	\begin{minipage}{\textwidth}
		%	\begin{adjustbox}{minipage=\linewidth,frame}
		\centering
		
		\begin{tabular}{@{}cc@{}}
			
			\includegraphics[clip, trim=0cm 0cm 0cm 0cm, scale=0.315]{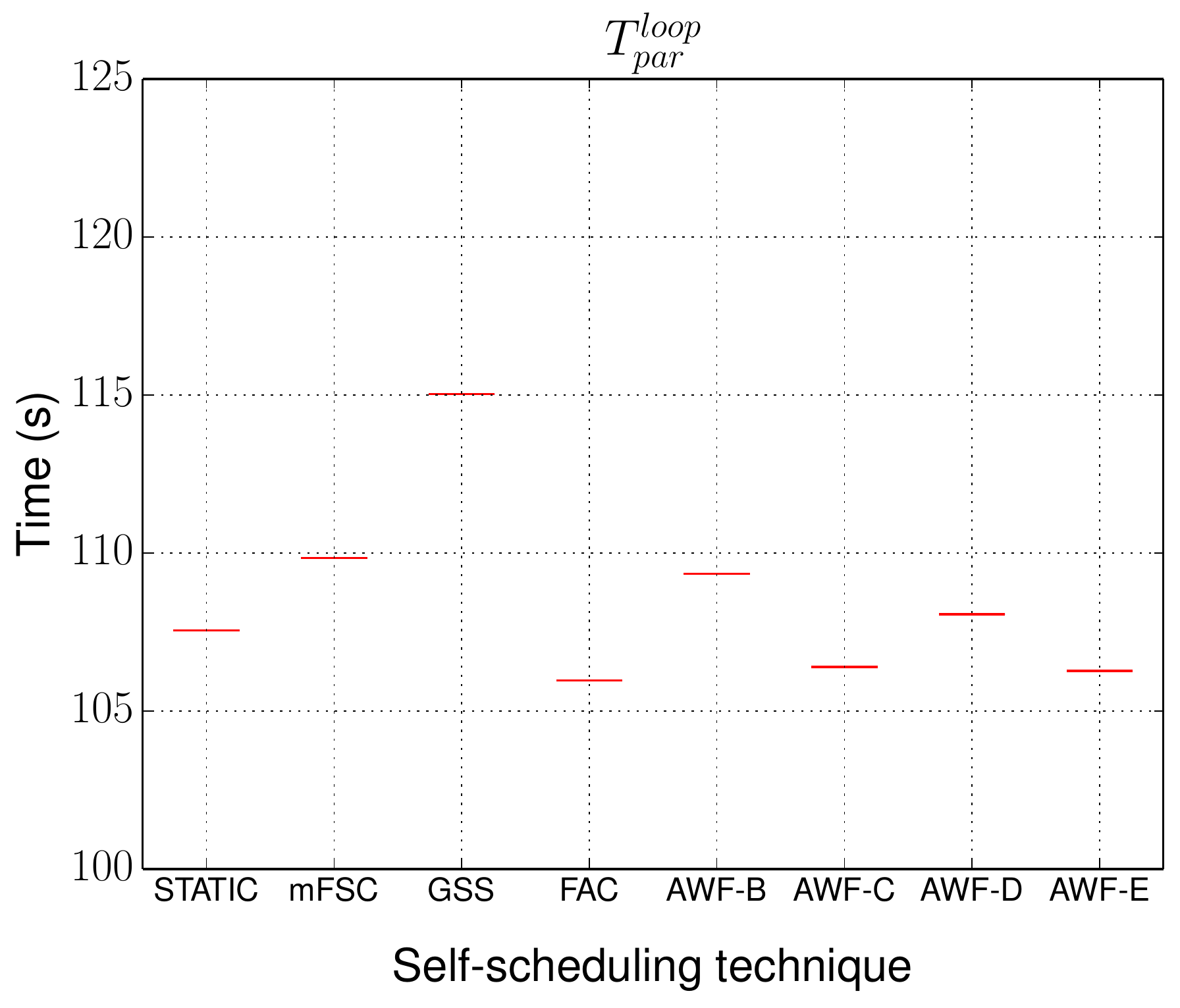}	& 	\includegraphics[clip, trim=0cm 0cm 0cm 0cm, scale=0.315]{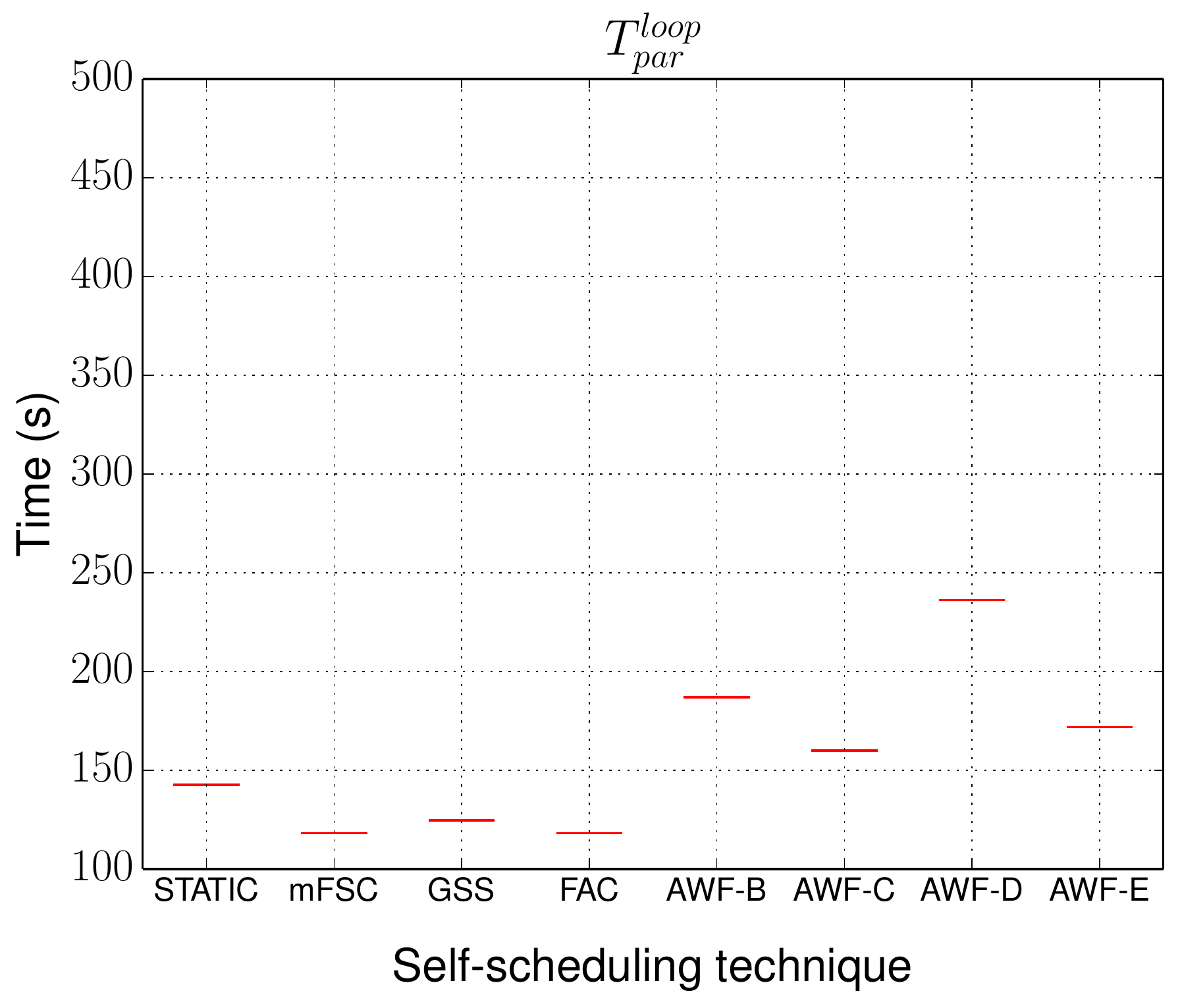} \\ 
			{\small (a) PSIA \alir{simulative} performance}	& (b) {\small Mandelbrot \alir{simulative} performance } \\ 
			
			\includegraphics[clip, trim=0cm 0cm 0cm 0cm, scale=0.315]{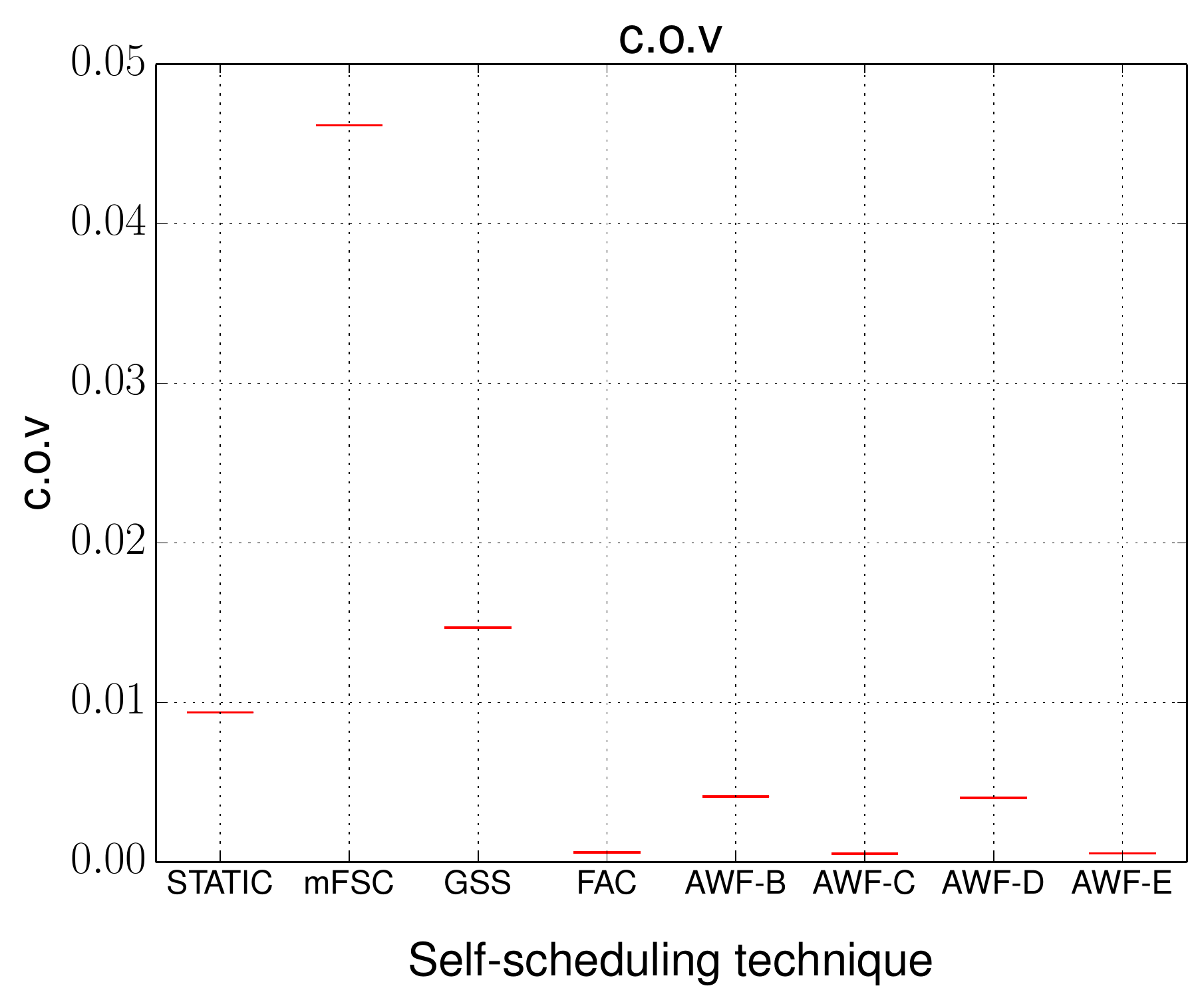}	& 	\includegraphics[clip, trim=0cm 0cm 0cm 0cm, scale=0.315]{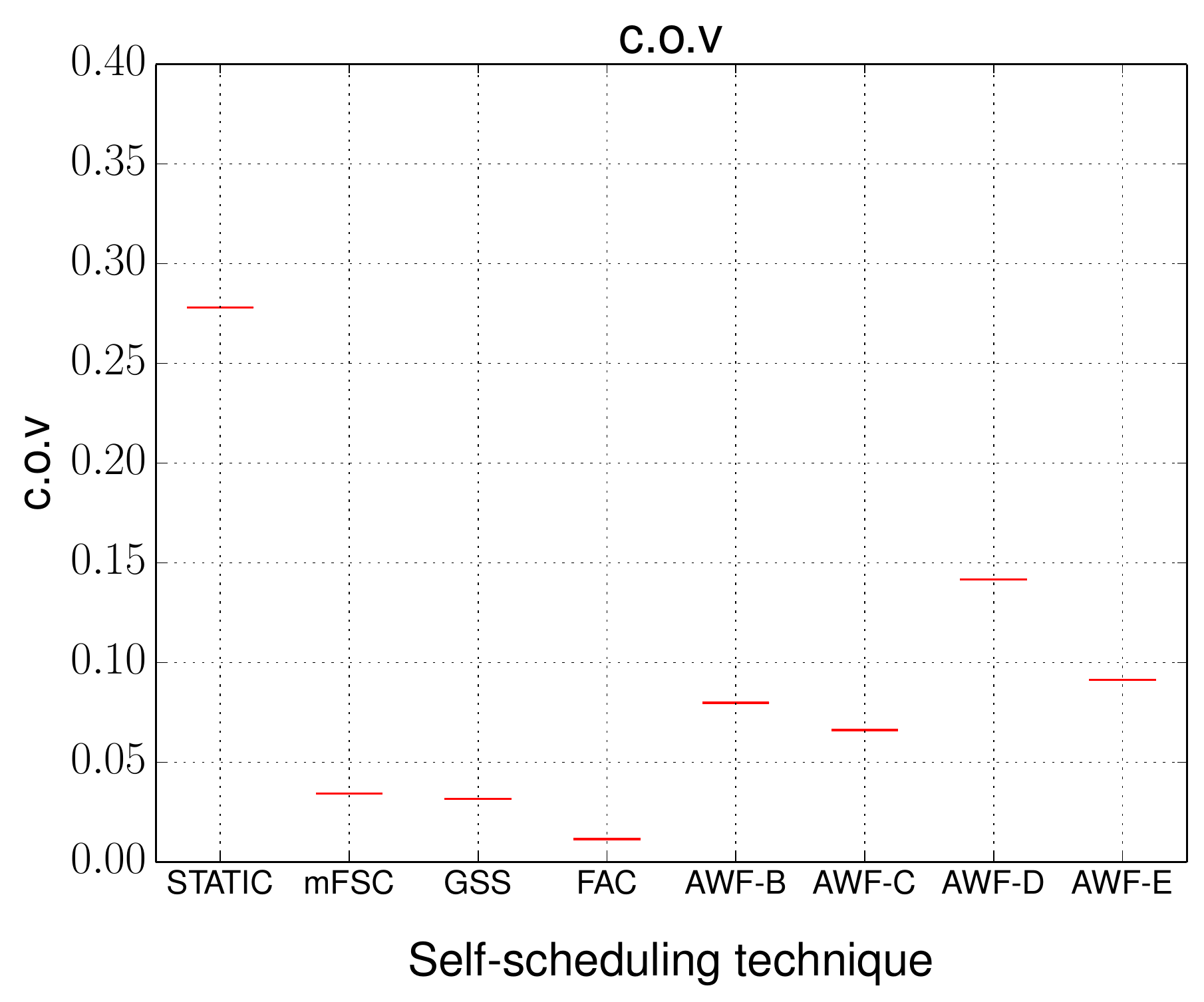} \\ 
			{\small(c) Load imbalance in PSIA (c.o.v.)}	& {\small(d)~Load imbalance in Mandelbrot (c.o.v.) } \\

			\includegraphics[clip, trim=0cm 0cm 0cm 0cm, scale=0.315]{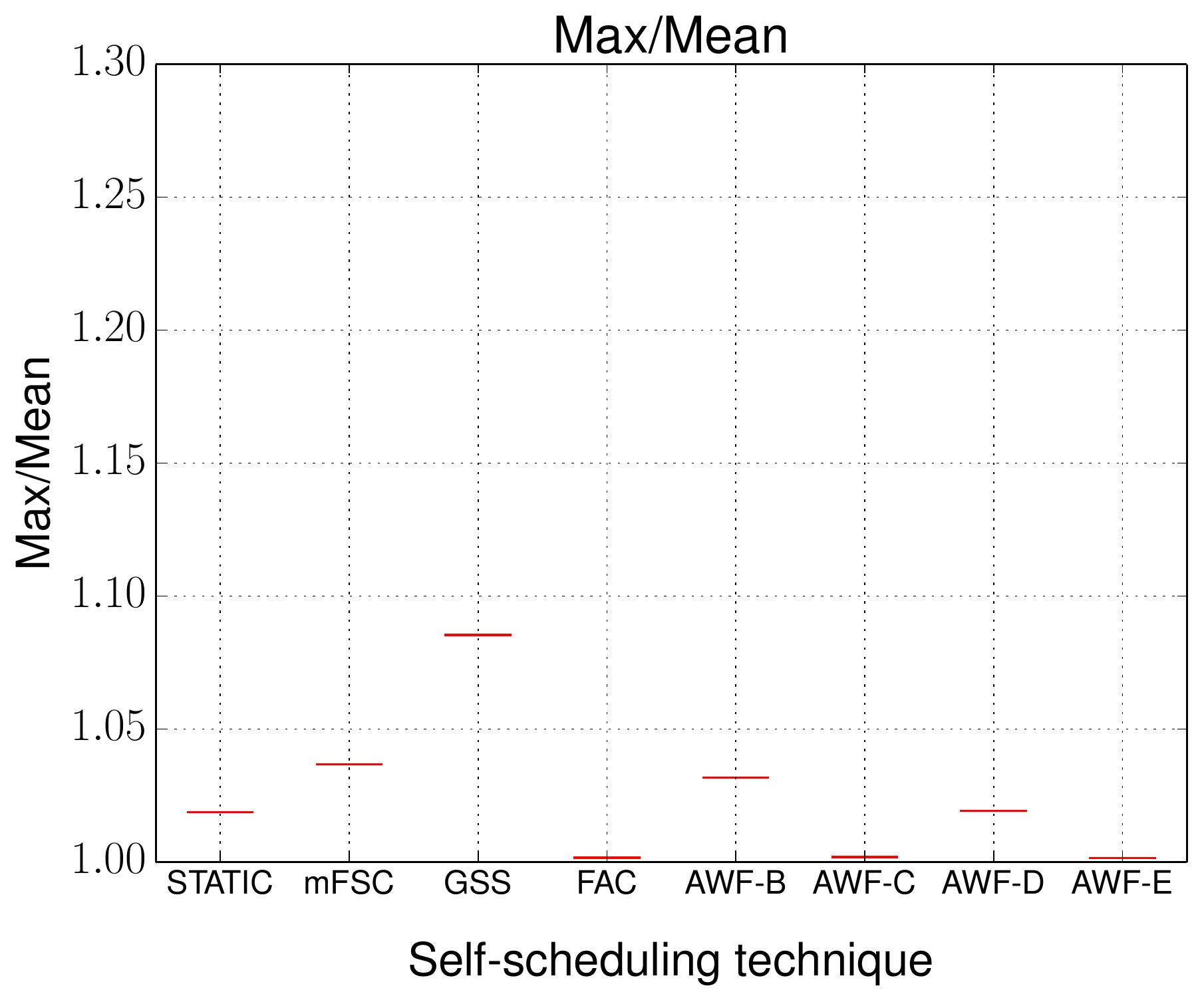}	& 	\includegraphics[clip, trim=0cm 0cm 0cm 0cm, scale=0.315]{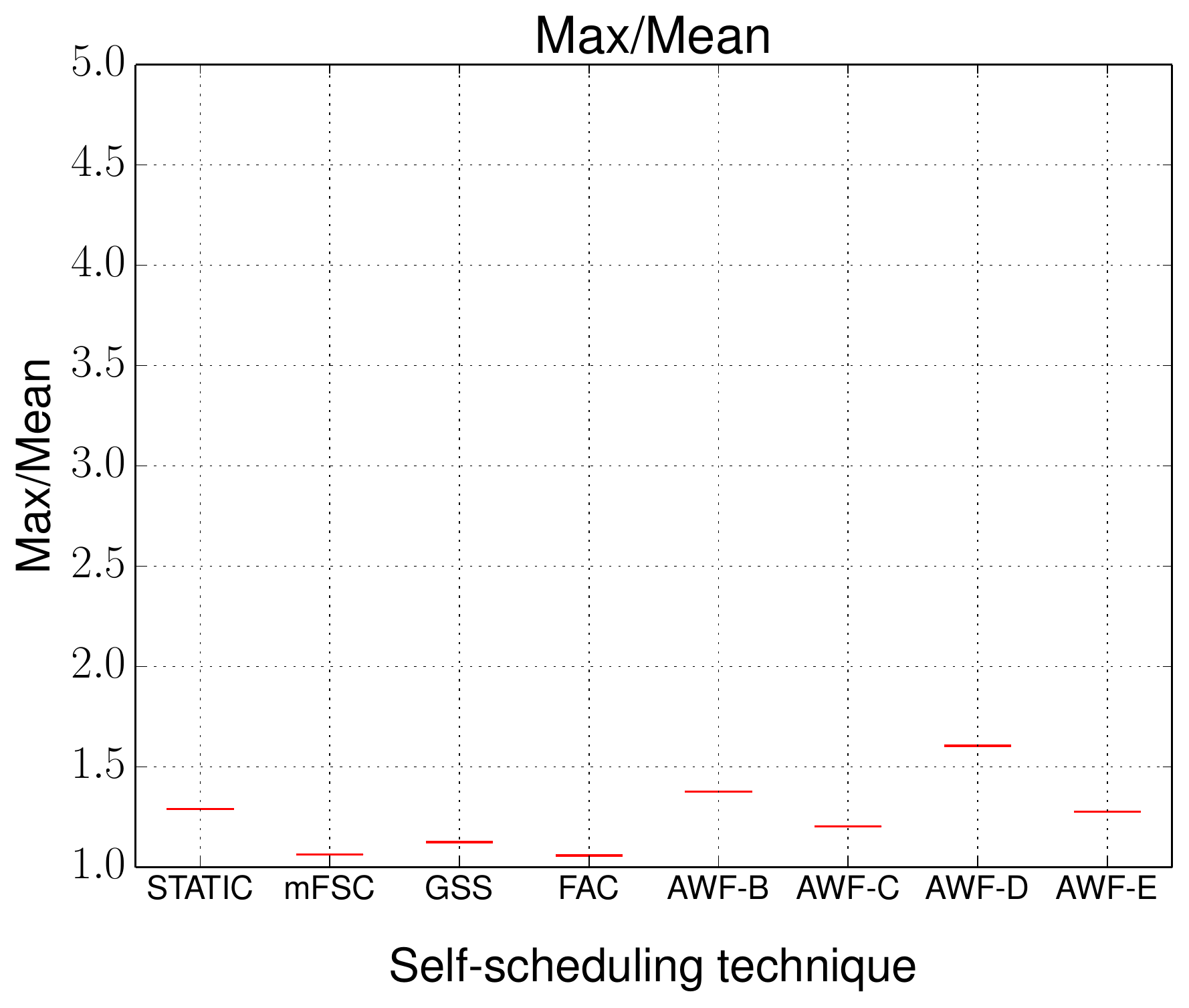} \\ 
			{\small (e) Load imbalance in PSIA (max/mean)} 	& {\small (f) Load imbalance in Mandelbrot (max/mean) }  \\ 
		\end{tabular}
	\end{minipage}
	\caption{\textbf{Simulative} performance of PSIA and Mandelbrot applications with reading \emph{FLOP\_file}. STATIC results in imbalanced load execution for PSIA and Mandelbrot and degrades the performance. GSS results in poor PSIA performance due to a process lagging the execution. FAC improves the performance of both application\fk{s} via a balanced load execution. Adaptive techniques result in enhanced PSIA performance and poor Mandelbrot performance. }
	\label{fig:simulative_flop_file}
	%	\vspace{-0.5cm}
\end{figure}

Two \fmc{application representation approaches are employed for the experiments using} \smpi{}+\msg{}.
The first \ali{approach} is denoted as \emph{FLOP\_file} and is shown in \figurename{~\ref{fig:simulative_flop_file}}.
The FLOP per task was measured with PAPI counters and was written into a file with task id and FLOP count per task.
This file is read by the simulator during the execution to account for the computational effort in each task.
Inspecting the first simulative performance results ( \emph{FLOP\_file}) in \figurename{~\ref{fig:simulative_flop_file}} reveals that STATIC degrades the performance of applications due to load imbalance as can be inferred from the load imbalance metrics in sub-figures(c-f). 
However, for STATIC with PSIA, the c.o.v and max/mean \ali{values are smaller} than that of mFSC and GSS.
\ali{The} GSS performance is worse than \ali{that of} mFSC, even though it has lower c.o.v. compared to mFSC for PSIA.
However, this is due to a single process lagging the execution of the PSIA as captured in sub-figure(e).
\ali{The} FAC technique results in improved performance for both applications.
The c.o.v. and max/mean \ali{values} with FAC in both applications is almost the minimum.
The adaptive techniques AWF-C and AWF-E improve the performance of PSIA and result in low parallel loop execution time, c.o.v., and max/mean almost similar to the FAC (the minimum).
AWF-B and AWF-D improve the performance of PSIA also, compared to mFSC and GSS. However, PSIA execution time with these techniques is slightly longer than the best (FAC, AWF-C, AWF-E).
The performance of Mandelbrot with the adaptive techniques is degraded in general compared to STATIC and dynamic nonadaptive DLS techniques.
This poor performance of Mandelbrot with adaptive techniques is due to the high load imbalance as indicated by the c.o.v. and max/mean metrics in sub-figures(d and f).
The high variability and the rather short execution time of the Mandelbrot left no room for the adaptive techniques to learn the correct relative PE weights.

\begin{figure}[]
	\begin{minipage}{\textwidth}
		%	\begin{adjustbox}{minipage=\linewidth,frame}
		\centering
		
		\begin{tabular}{@{}cc@{}}
			
			\includegraphics[clip, trim=0cm 0cm 0cm 0cm, scale=0.315]{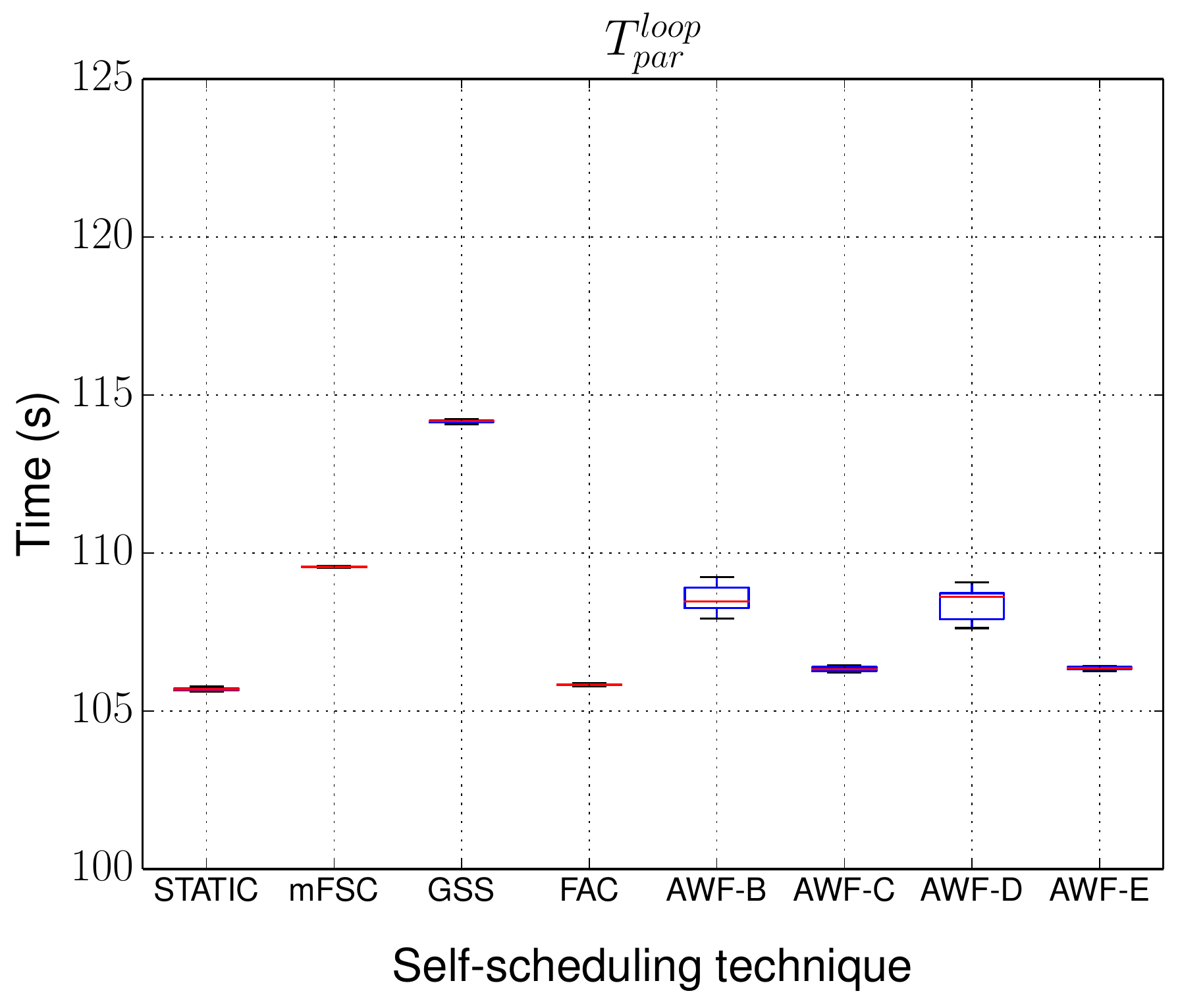}	& 	\includegraphics[clip, trim=0cm 0cm 0cm 0cm, scale=0.315]{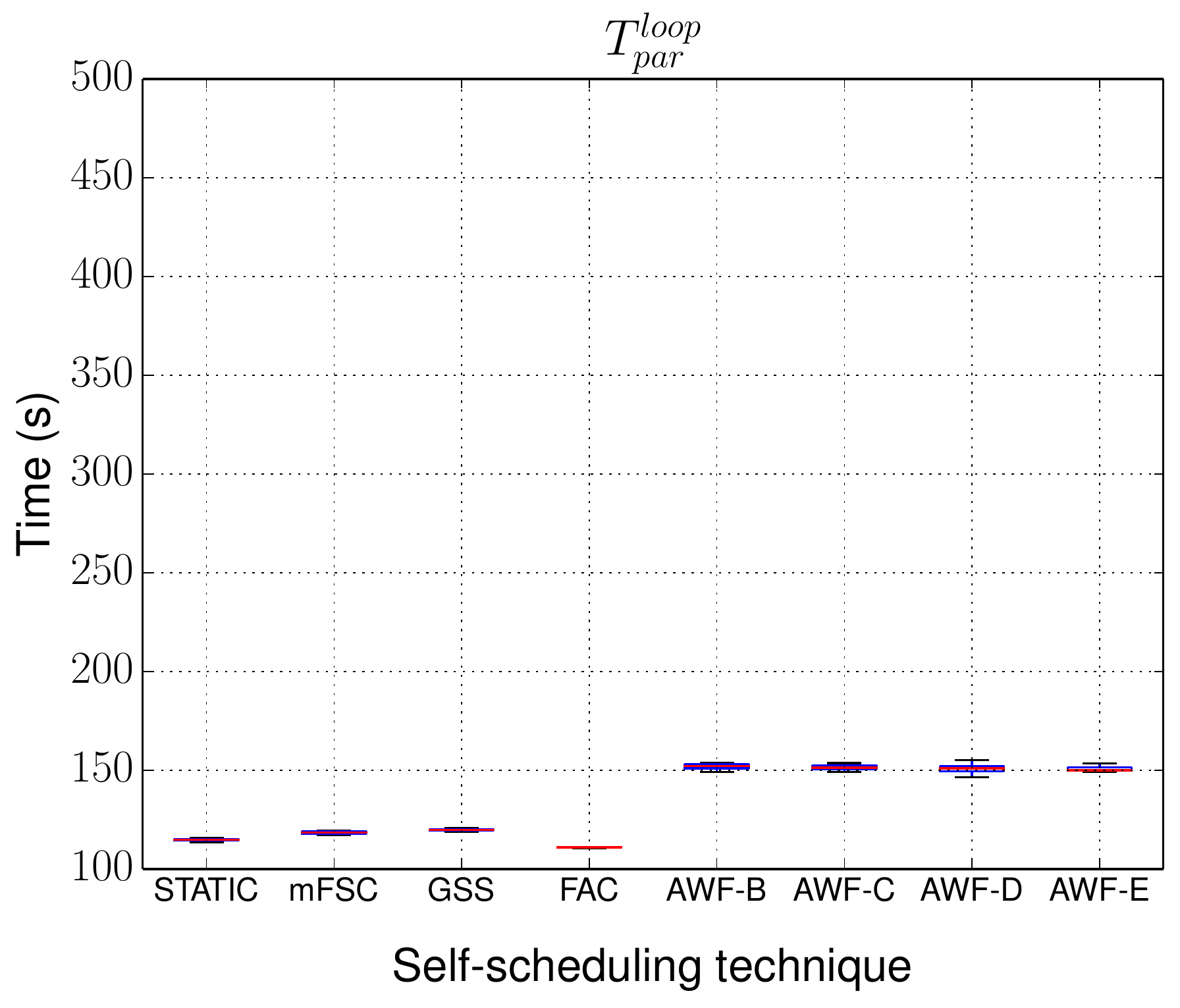} \\ 
			{\small (a) PSIA \alir{simulative} performance}	& (b) {\small Mandelbrot \alir{simulative} performance } \\ 
			
			\includegraphics[clip, trim=0cm 0cm 0cm 0cm, scale=0.315]{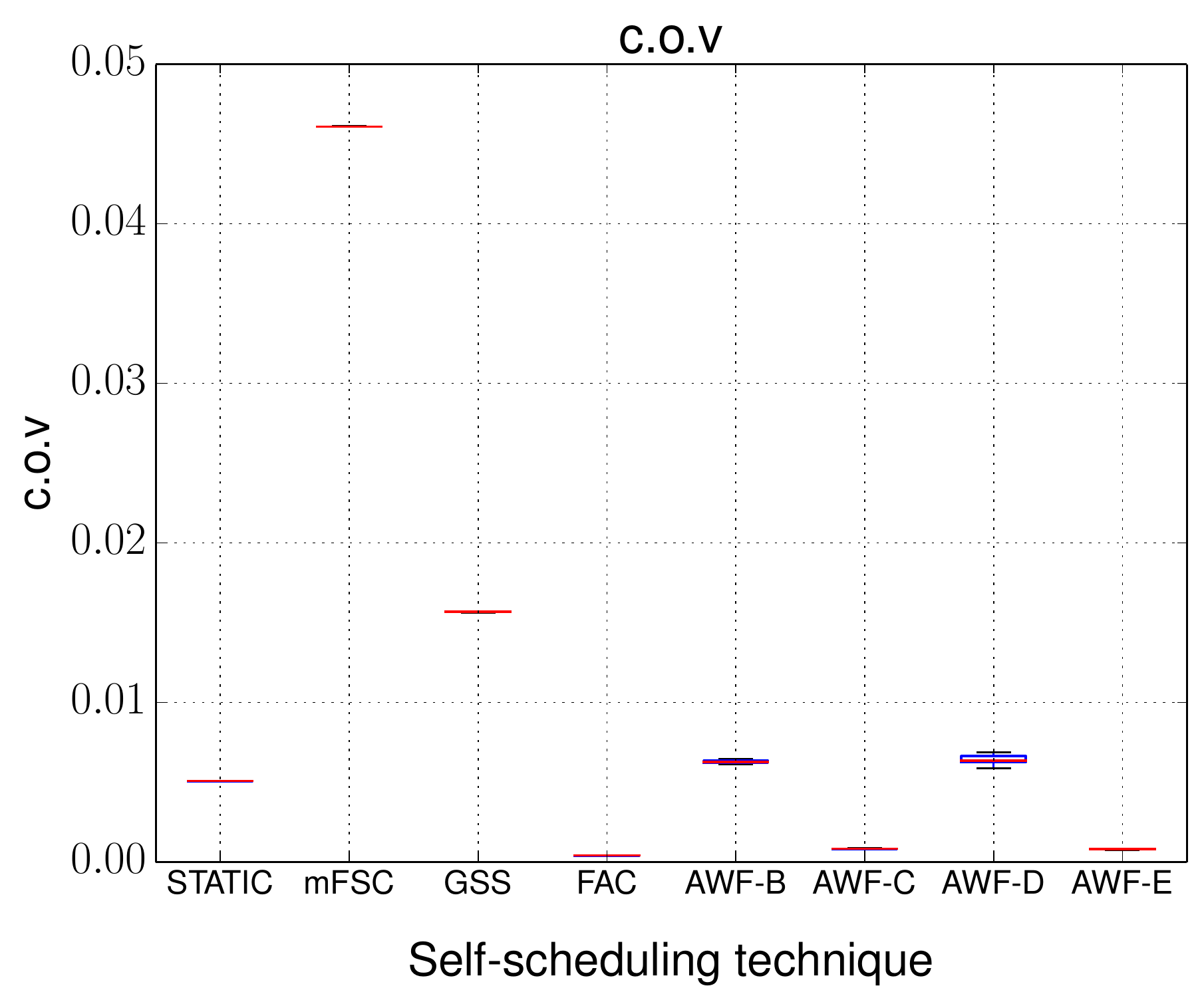}	& 	\includegraphics[clip, trim=0cm 0cm 0cm 0cm, scale=0.315]{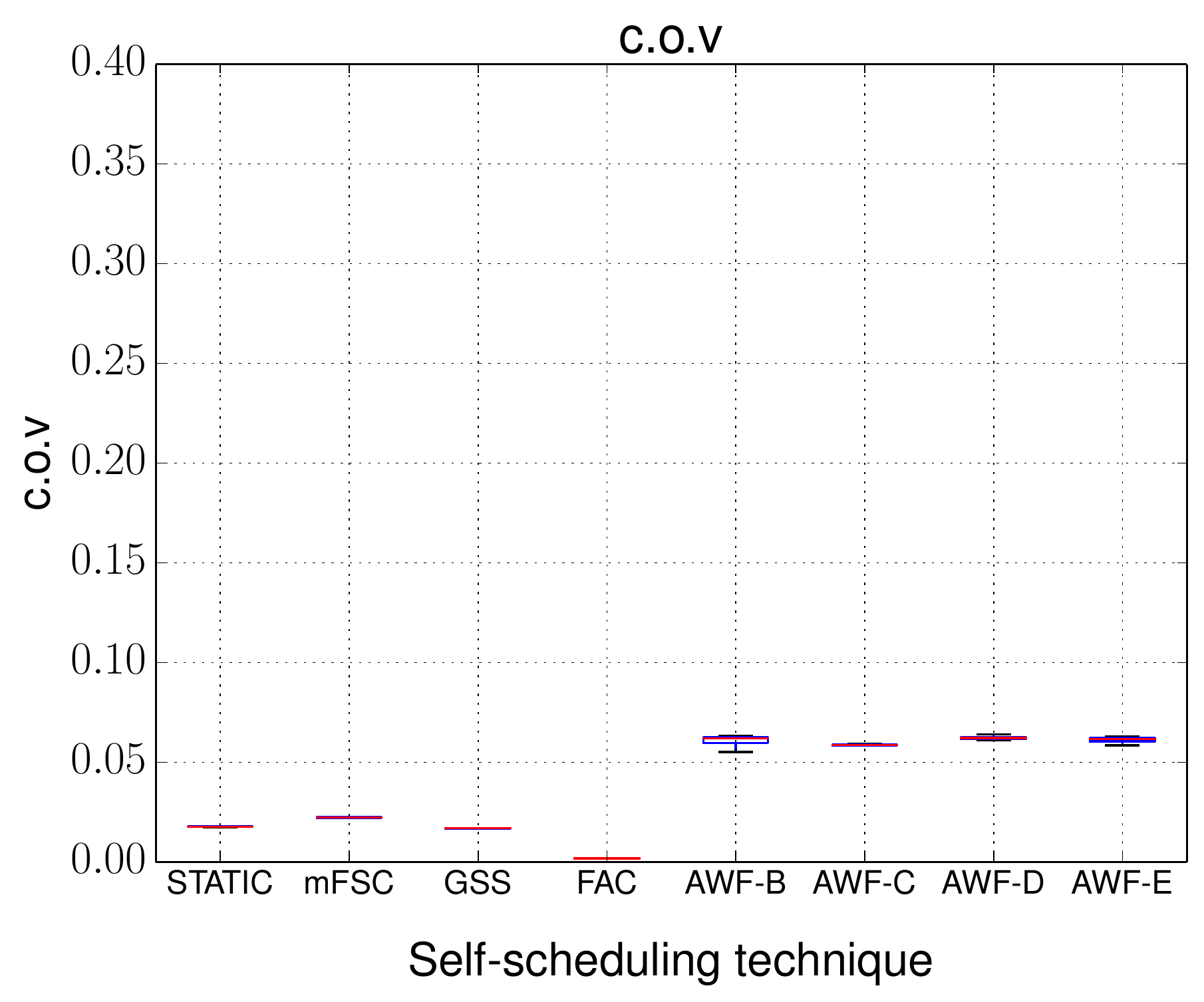} \\ 
			{\small(c) Load imbalance in PSIA (c.o.v.)}	& {\small(d)~Load imbalance in Mandelbrot (c.o.v.) } \\

			\includegraphics[clip, trim=0cm 0cm 0cm 0cm, scale=0.315]{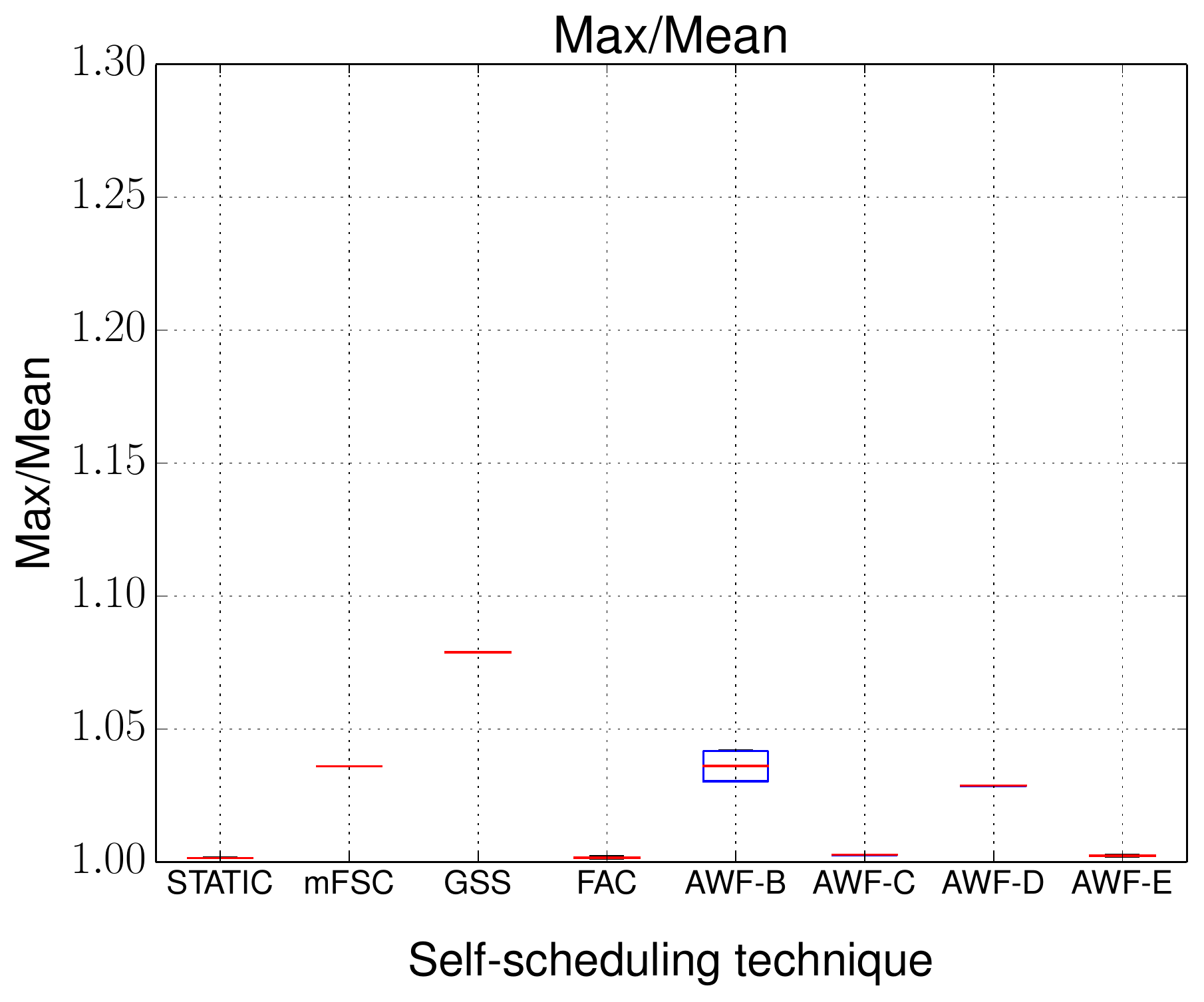}	& 	\includegraphics[clip, trim=0cm 0cm 0cm 0cm, scale=0.315]{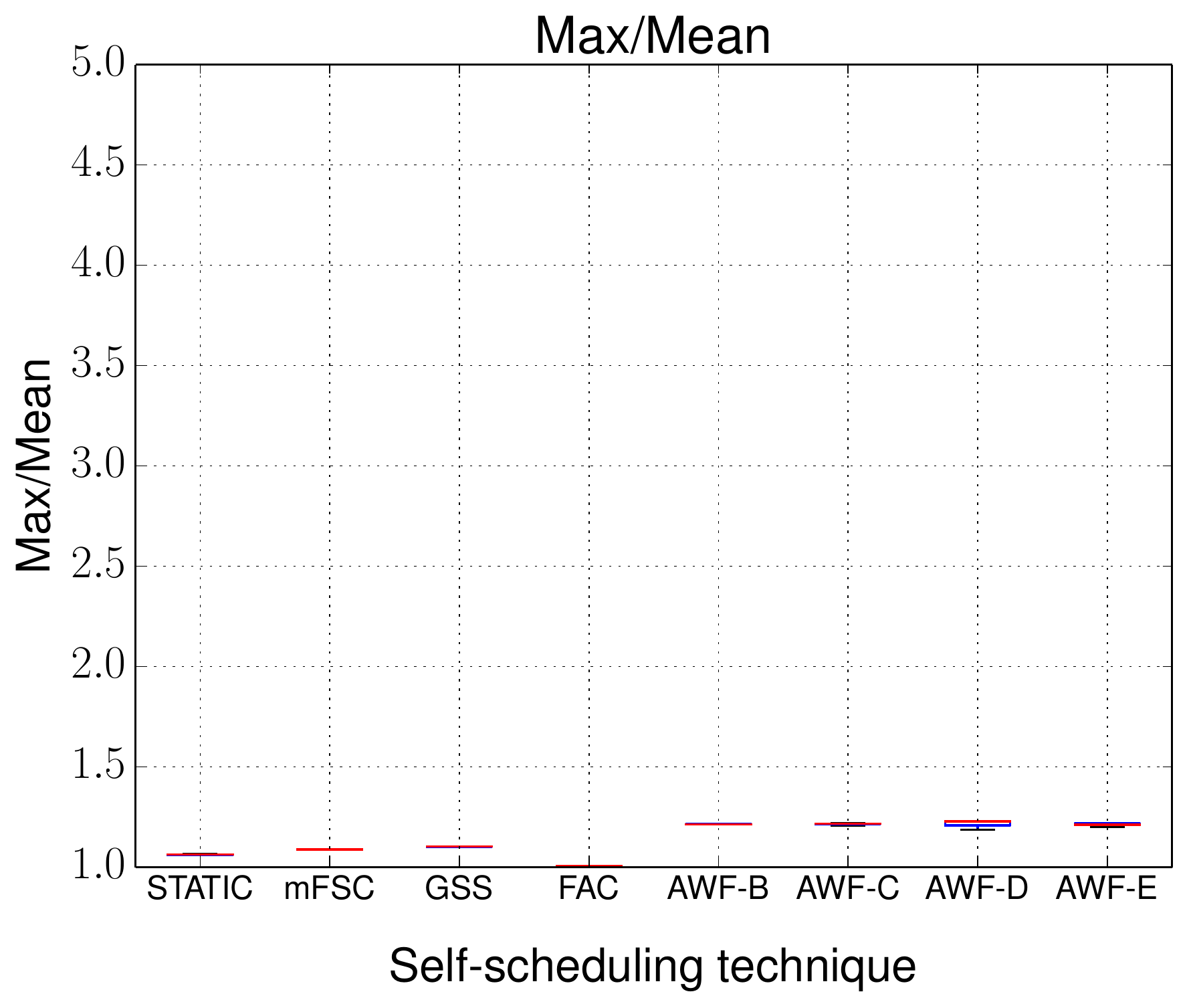} \\ 
			{\small (e) Load imbalance in PSIA (max/mean)} 	& {\small (f) Load imbalance in Mandelbrot (max/mean) }  \\ 
		\end{tabular}
	\end{minipage}
	\caption{\textbf{Simulative} performance of PSIA and Mandelbrot applications with \emph{FLOP distribution}. STATIC, FAC, AWF-C, AWF-E results in the best PSIA performance. GSS degrades PSIA performance and mFSC results in high load imbalance. FAC achieves the best performance for both applications. Adaptive techniques degrade Mandelbrot performance.}
	\label{fig:simulative_flop_dist}
	%	\vspace{-0.5cm}
\end{figure}
The second \ali{simulation approach is denoted as} \emph{FLOP\_dist} and is shown in \figurename{~\ref{fig:simulative_flop_dist}}.
The measured FLOP counts with PAPI is used to fit a probability distribution to the measured FLOP data as described in Section~\ref{subsec:realsim} above.
In this case, the simulation is repeated $20$ times similar to the native execution with different seeds to capture the variability of the performance of the native application.
Inspecting the first simulative performance results (FLOP dist.) in \figurename{~\ref{fig:simulative_flop_dist}} reveals that applications performance with STATIC is better than mFSC and GSS techniques, and almost similar to the best performance achieved by FAC. 
This is assured by the low values of the load imbalance metrics for both applications with STATIC.
GSS degrades the PSIA performance due a process lags the application execution time as indicated by a high max/mean value.
mFSC also failed to balance the load of PSIA as indicated by a high c.o.v. value and results in a long parallel loop execution time.
FAC results in the shortest parallel loop execution times, c.o.v. and max/mean values for both applications under test. 
Adaptive techniques in general, and specifically AWF-C and AWF-E improve the PSIA performance with balanced load execution and results in the shortest execution time (similar to FAC).
However, adaptive techniques failed to adapt correctly to the high variability of tasks execution times of Mandelbrot, due to its short execution time and resulted in poor performance.

\begin{figure}[]
	\begin{minipage}{\textwidth}
		%	\begin{adjustbox}{minipage=\linewidth,frame}
		\centering
		
		\begin{tabular}{@{}cc@{}}
			
			\includegraphics[clip, trim=0cm 0cm 0cm 0cm, scale=0.315]{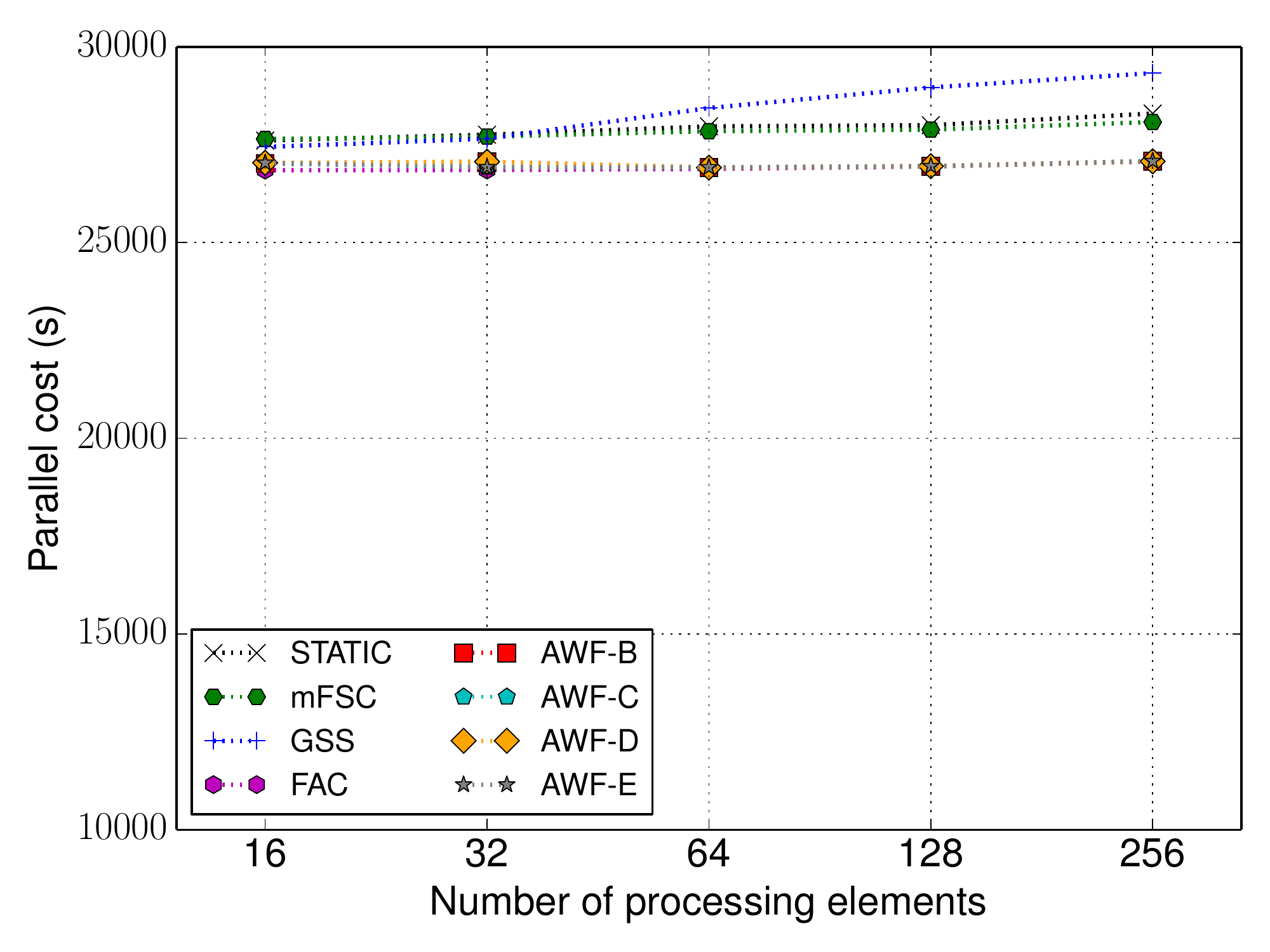}	& 	\includegraphics[clip, trim=0cm 0cm 0cm 0cm, scale=0.315]{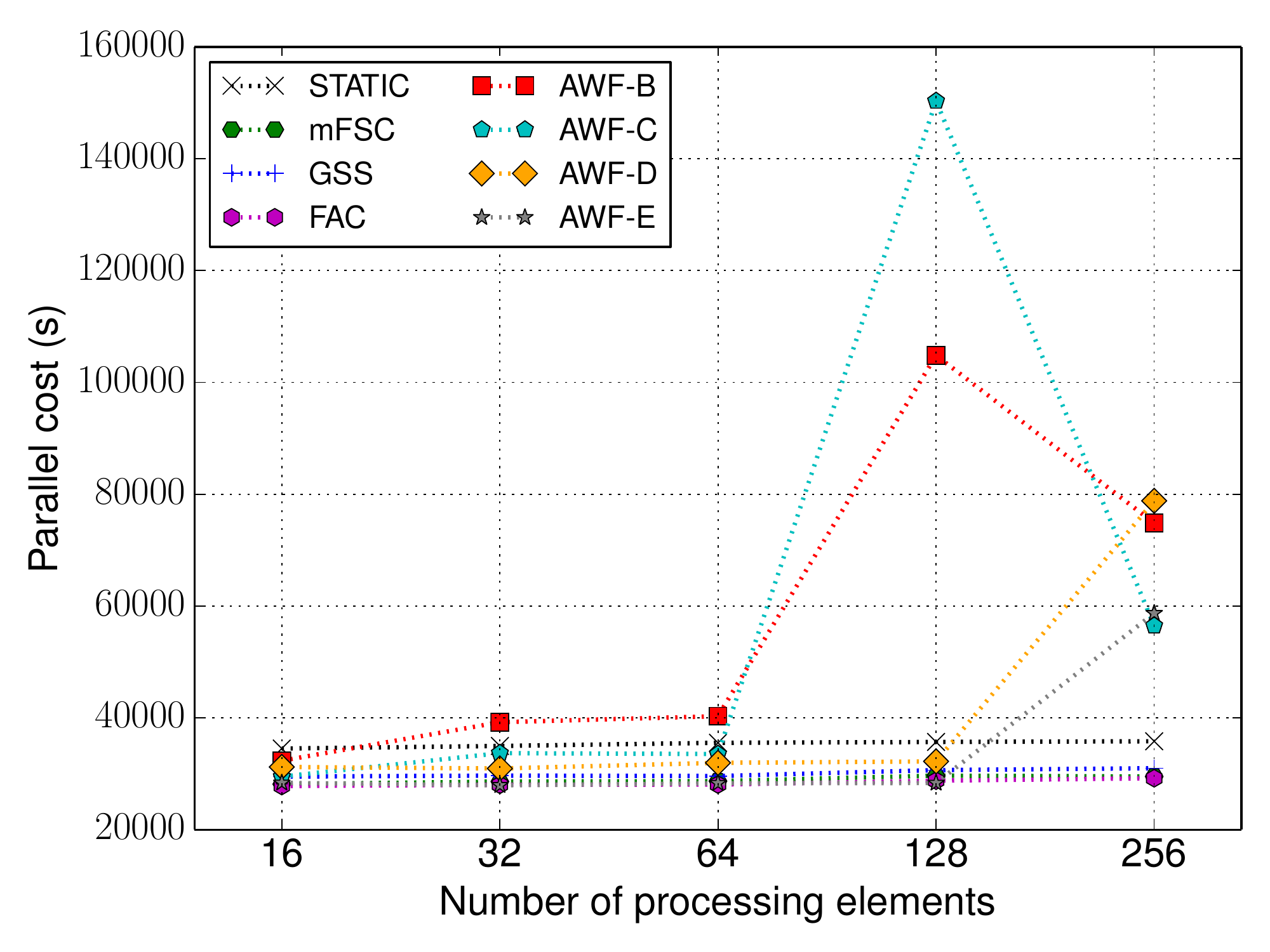} \\ 
			{\small (a) PSIA strong scaling}	& (b) {\small Mandelbrot strong scaling } \\ 
			
			\includegraphics[clip, trim=0cm 0cm 0cm 0cm, scale=0.315]{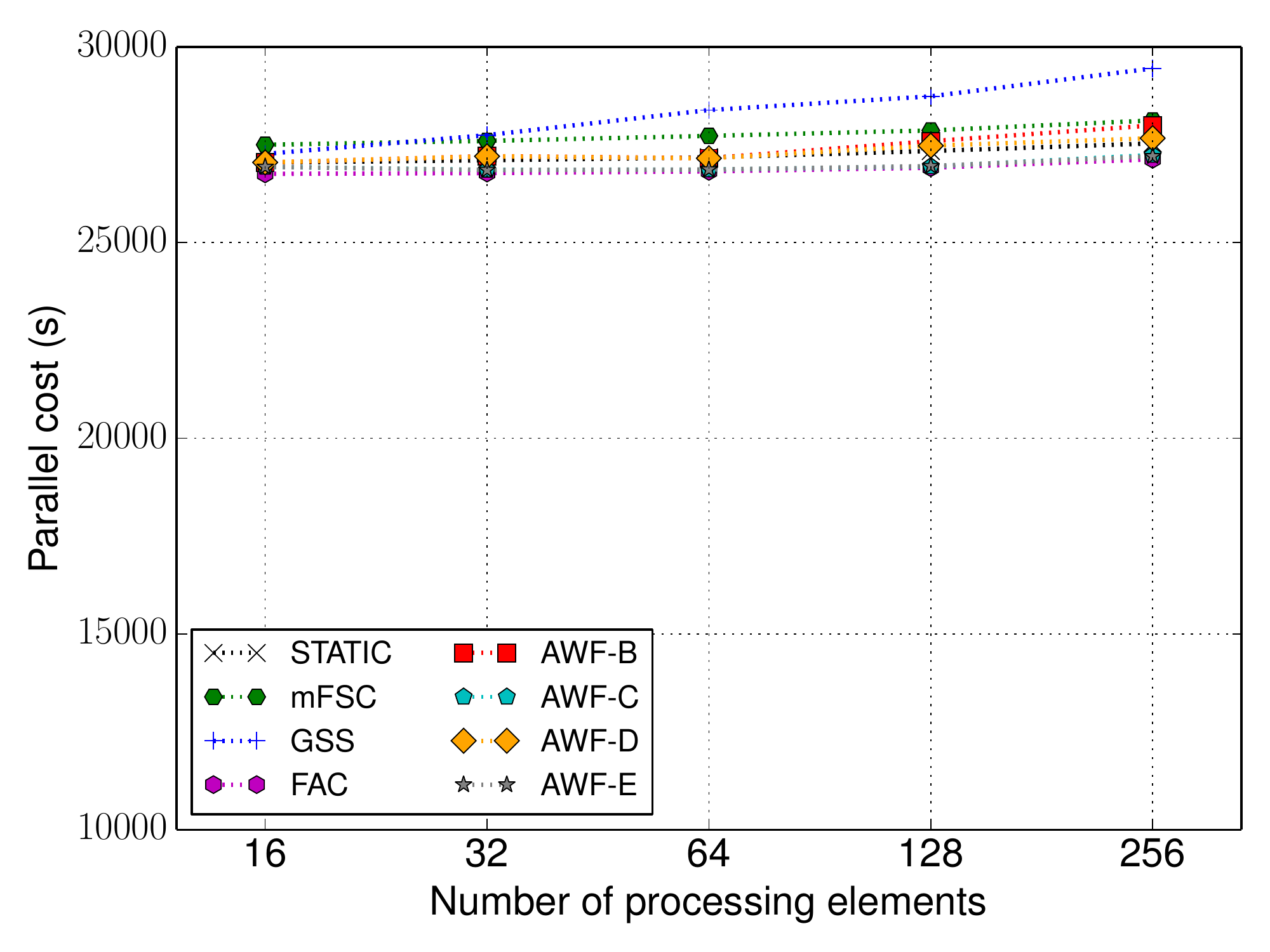}	& 	\includegraphics[clip, trim=0cm 0cm 0cm 0cm, scale=0.315]{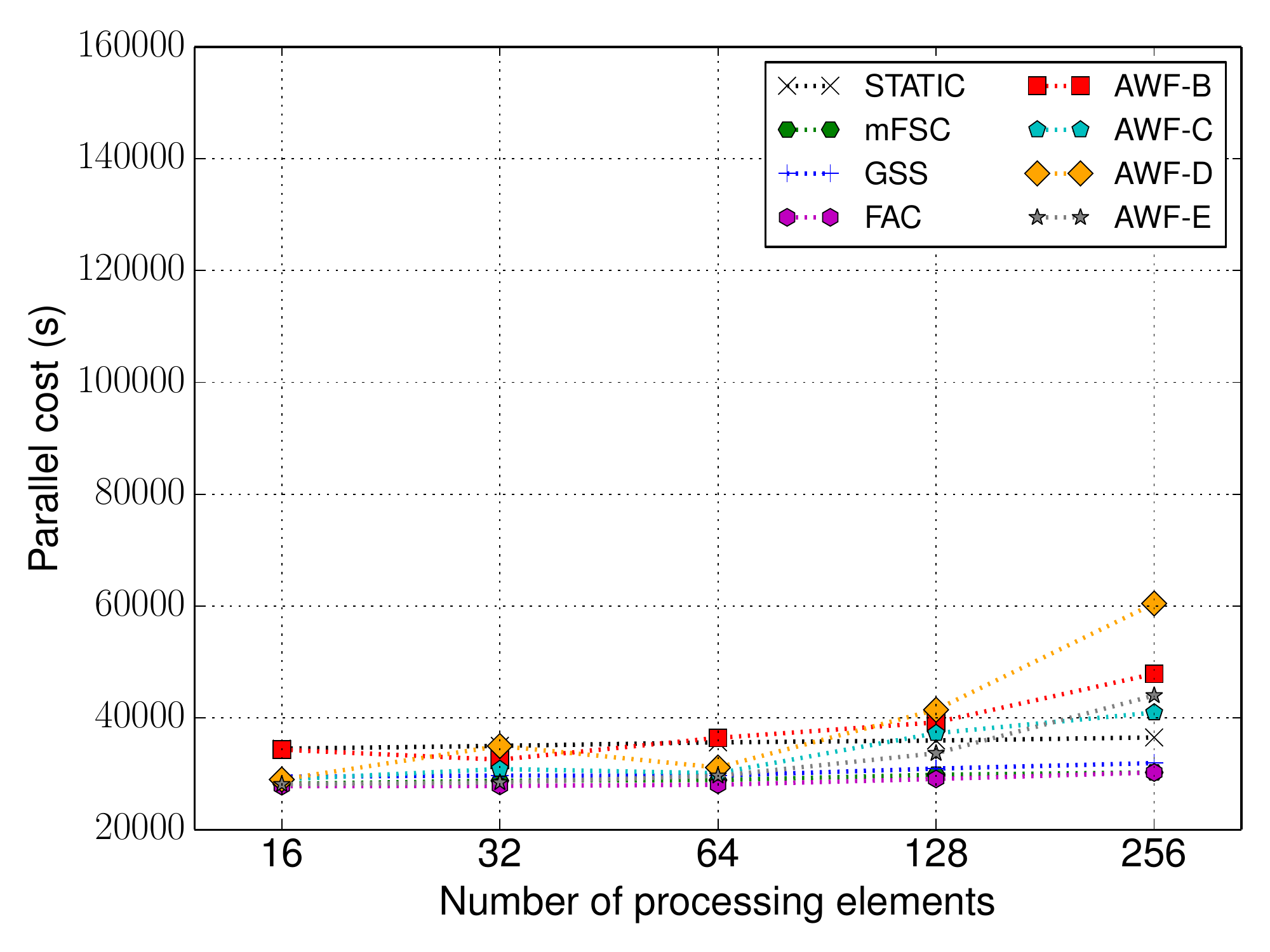} \\ 
			{\small(c) PSIA strong scaling (\emph{FLOP\_file})}	& {\small(d)~Mandelbrot strong scaling (\emph{FLOP\_file}) } \\

			\includegraphics[clip, trim=0cm 0cm 0cm 0cm, scale=0.315]{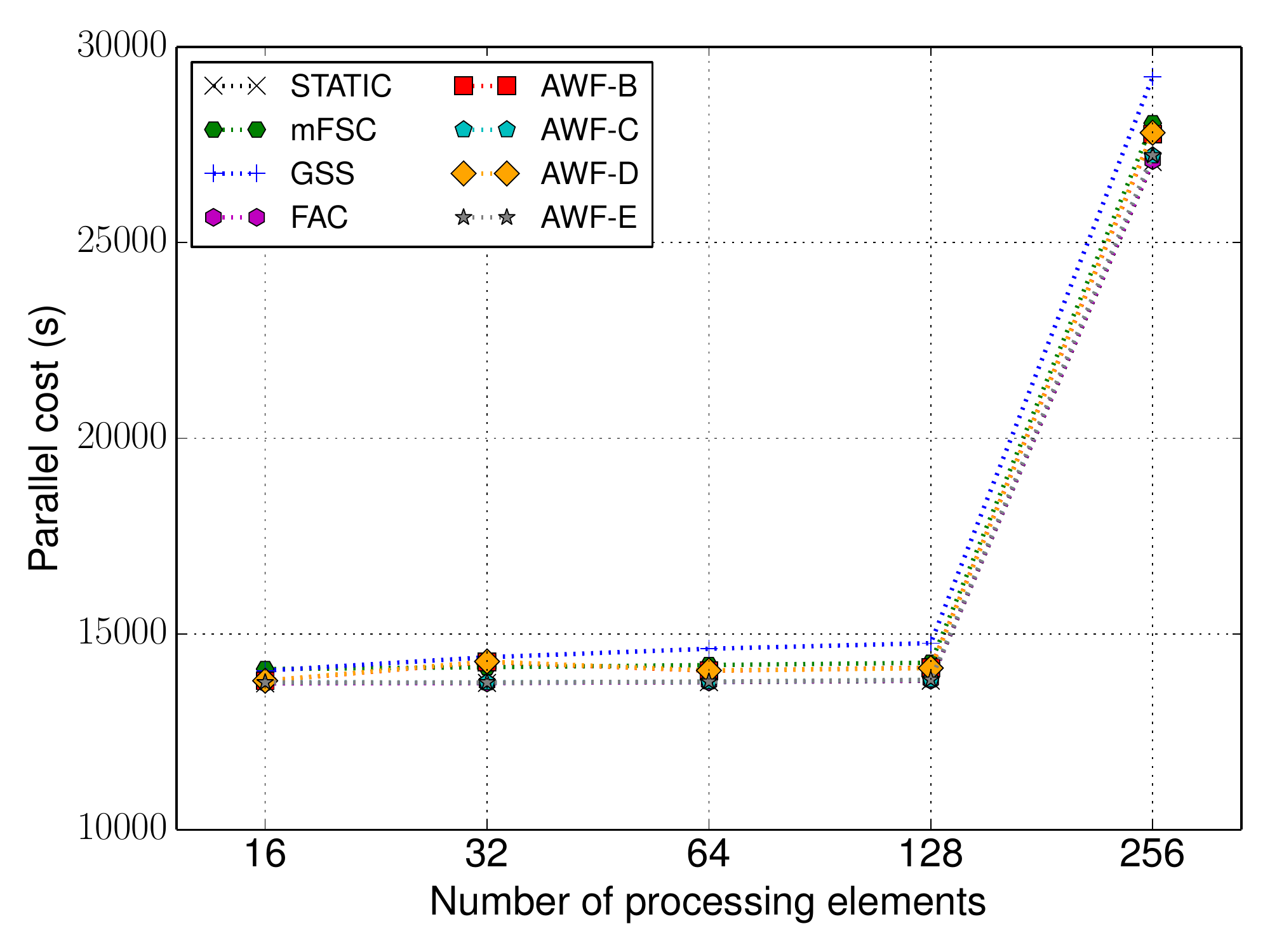}	& 	\includegraphics[clip, trim=0cm 0cm 0cm 0cm, scale=0.315]{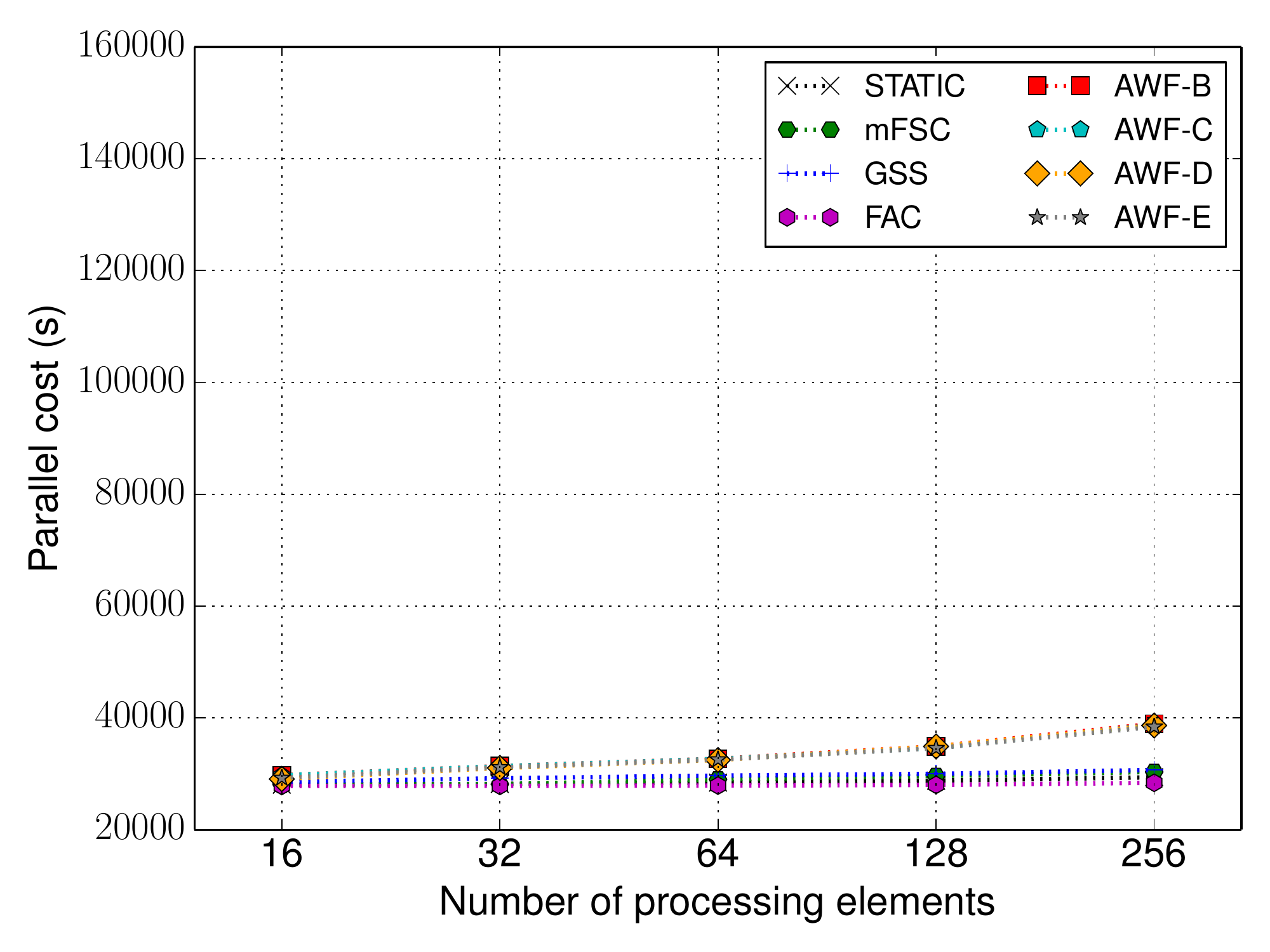} \\ 
			{\small (e) PSIA strong scaling (\emph{FLOP\_dist.})} 	& {\small (f) Mandelbrot strong scaling (\emph{FLOP\_dist.}) }  \\ 
		\end{tabular}
	\end{minipage}
    \caption{Strong scaling of \textbf{native} (subfigures (a) and (b)) and  \textbf{simulative} (subfigures (c) - (f)) performance of PSIA and Mandelbrot applications. The simulative results are shown for the first simulation approach, \emph{FLOP\_file}, as well as, for the second one, \emph{FLOP\_dist}.}
	\label{fig:scaling}
	%	\vspace{-0.5cm}
\end{figure}

\alir{\subsubsection{Strong scaling results}}
\fk{In \figurename{~\ref{fig:scaling}}, the \fmc{strong} scaling behavior of the PSIA and Mandelbrot applications is shown for the native (subfigures (a) and (b)) and simulative executions (subfigures (c) - (f)), respectively.
Considering the native executions of PSIA, all DLS techniques scale very well. 
FAC and the adaptive \fmc{DLS} techniques show a constant parallel cost, while the parallel cost increases slightly with the increasing number of processing elements for mFSC and STATIC.
The largest slope is induced by the execution using the GSS technique.
By contrast, an almost constant parallel cost of the Mandelbrot performance is \fmc{obtained with} mFSC, GSS, and FAC.
The parallel cost of using STATIC is also almost constant but higher than that of using \fmc{mFSC, GSS, and FAC}.
Using the adaptive DLS techniques results in \fmc{poorer} strong scaling, characterized by at least one outlier per adaptive technique.}

\fk{The strong scaling results for the first simulation approach, denoted as \emph{FLOP file}, are shown in 
\figurename{~\ref{fig:scaling} (c)-(d)} for the PSIA and Mandelbrot applications, respectively.
While the parallel costs are almost equal to the parallel costs of the native executions of PSIA, this is not the case for the Mandelbrot application.
The Mandelbrot simulations show almost constant parallel costs for mFSC, GSS, FAC, and STATIC.
\fmc{These results are identical to those} of the native executions.
Considering the adaptive \fmc{DLS} techniques, the parallel costs are not characterized by outliers as observed for the native executions.
However, in contrast to the non-adaptive DLS \fmc{techniques}, the parallel cost increases with the number of processing elements.}

\fk{Considering the simulative executions using the second simulation approach, denoted as \emph{FLOP dist.}, a \fmc{rather} different \fmc{strong} scaling behavior is observed.
For the PSIA application, the parallel costs are equal to the parallel costs of the native executions only for $256$ processing elements. 
For a lower number of processing elements, the parallel costs are approximately half of those of the native executions.
The parallel costs of the simulative executions of the Mandelbrot application are almost constant.
Only using the adaptive \fmc{DLS} techniques results in a slight increase of the parallel costs with increasing number of processing elements. 
}

\subsection{Discussion}
\label{subsec:discussion}

To evaluate how realistic the performed simulations are, the native and simulative performance of PSIA and Mandelbrot is analyzed in terms of $T^{loop}_{par}$, c.o.v., and max/mean \ali{metrics}.
Realistic simulation results \ali{are expected to} lead to a similar analysis and \ali{similar} conclusions \ali{drawn from} the analysis of the native results.

Table~\ref{tbl:realistic} summarizes seven performance features form the performance analysis of applications' performance with various scheduling techniques performed above in Section~\ref{subsec:results}. 
\ali{The comparison between the native and simulative} performance analysis results shows that \ali{the} simulations with \emph{FLOP file} captured almost all the performance features that characterize the performance of the two applications under test. 
The simulator overestimated only the performance of AWF-B and AWF-D.

% \usepackage{booktabs}
\begin{table}[!h]
	\centering
	\caption{Native application performance features \ali{realistically} captured by simulations}
	\label{tbl:realistic}
	\begin{adjustbox}{max width=\textwidth}
		\begin{tabular}{@{}lll@{}}
			\toprule
			\textbf{Performance features}                      & \textbf{\begin{tabular}[c]{@{}l@{}}SMPI+MSG\\ FLOP file\end{tabular}} & \textbf{\begin{tabular}[c]{@{}l@{}}SMPI+MSG\\ FLOP\_dist\end{tabular}} \\ \midrule
			Load imbalance with STATIC (PSIA, Mandelbrot)                  & Captured                                                              & Not captured                                                           \\
			High c.o.v. with mFSC (PSIA)                       & Captured                                                              & Captured                                                               \\
			Long $T_{par}^{loop}$, low c.o.v., and max/max with GSS (PSIA) & Captured                                                              & Captured                                                               \\
			FAC best performance (PSIA, Mandelbrot)                        & Captured                                                              & Captured                                                               \\
			Adaptive techniques high performance (PSIA)        & Partially captured                                                    & Partially captured                                                     \\
			Adaptive techniques poor performance (Mandelbrot)  & Captured                                                              & Captured                                                               \\ 
			Adaptive techniques high variability (Mandelbrot)  & Not captured                                                              & Not captured        \\
			\midrule
			\multicolumn{3}{c}{\alir{\textbf{Strong scaling experiments}}}  \\ \midrule
			\alir{mFSC and STATIC slight increase in parallel cost (PSIA)} & \alir{Captured}  & \alir{Not captured} \\
			\alir{FAC and adaptive techniques constant parallel cost (PSIA)} &  \alir{  Captured }  & \alir{Not captured} \\
			\alir{GSS poor scalability (PSIA)} &  \alir{Captured} & \alir{Not captured} \\
			\alir{STATIC constant and high parallel cost (Mandelbrot)} & \alir{ Captured }& \alir{Captured} \\
			\alir{mFSC, GSS, and FAC almost constant cost (Mandelbrot)} & \alir{ Captured} & \alir{Captured} \\
			\alir{Adaptive techniques poor scaling and outliers (Mandelbrot)} & \alir{Partially captured} & \alir{Partially captured} \\
			\bottomrule
		\end{tabular}
	\end{adjustbox}
\end{table}

Both simulations predicted correctly that \ali{the} FAC technique achieves a balanced load execution for both applications and improves performance.
Simulations with \ali{the} \emph{FLOP\_dist} \fmc{approach} failed to capture the load imbalance with STATIC in both applications.
The performance with STATIC is significantly affected by the order of tasks or loop iterations assigned to each PE. 
As the order of tasks is not preserved by drawing random samples from FLOP distributions, the load imbalance with STATIC was dissolved between PEs as they are assigned different tasks in simulative execution from the native one.
Interestingly, both simulations were able the most devious performance feature of high $T_{par}^{loop}$, low c.o.v, and high max/mean \ali{values} of GSS with PSIA.
Both simulations did not capture the high variability of adaptive techniques. %as this high variability is an artifact of the probing effect, which does not exist in the simulative execution.
\ali{The} adaptive techniques depend on time measurements to estimate PE performance.
If the granularity of the tasks is highly variable, and some task sizes are very fine, the time measurement of their execution will be inaccurate due to an overhead of the time measurement.
The inaccurate time measurement leads to incorrect weight estimation and high variability between different native executions.
This probing effect does not exist in the simulative execution and, therefore, was not fully captured.
However, both simulations \ali{correctly} predicted the high performance of adaptive techniques with PSIA and their low performance with Mandelbrot.
The simulation with \emph{FLOP\_dist} was able to capture the small variability in performance with various DLS techniques, \ali{which} was not captured by reading the FLOP counts from a file in the first simulation.

%\begin{itemize}
%	\item general comments on the results
%	\item MSG vs SMPI+MSG
%	\item  what is realistic - definition - challenges 
%\end{itemize}
%

% !TEX root =  fgcs_19.tex
\section{Conclusion and Future Work} 
\label{sec:conc}
%\begin{enumerate}
%	\item Discuss and reiterate major advantages and drawbacks of the new solution
%	\item List future work that can be done using the results of the current article
%	\item[Future work]
%	\item Decentralized \dlbTool{} implement, simulate, verify
%	\item More simulators, systems
%	\item Automate SMPI+MSG or SMPI+S4U as a tool in the future
%\end{enumerate}
\ahmed{
In this work, we show that it is possible to \emph{realistically simulate the performance of scientific applications on HPC systems}.
The approach proposed \fmc{for this purpose} considers various factors that affect \fmc{the} applications performance on HPC systems: application representation, scheduling, computing system representation, and \emph{systemic} variations.
The proposed realistic simulation approach has been exemplified \fmc{on} two \mbox{computationally-intensive} scientific applications.
A set of guidelines are \fmc{also} introduced and discussed \fmc{for how} to represent applications and system characteristics.
These guidelines help to achieve realistic simulations irrespective of the application type (e.g., communication- or \mbox{computationally-intensive}) and the simulation toolkit (e.g., Alea or \mbox{GridSim}~\cite{klusavcek2010alea}).
}

\ahmed{Based on the proposed approach,} a novel simulation \ahmed{method} \alir{is also introduced} for the accurate and fast simulation of \mbox{MPI-based} applications.
\ahmed{This method \fmc{jointly} employs \fmc{SimGrid's} SMPI+MSG \fmc{interfaces} to simulate applications performance with minimal changes \fmc{to the original application source code}.
We used this method to realistically simulate  two \mbox{computationally-intensive} scientific applications using different scheduling techniques.}
The comparison of performance characteristics extracted from \ahmed{the} native and simulative \ahmed{results} shows that the proposed simulation \fk{approach} captured \fmc{very closely} most of the performance characteristics of interest\ahmed{, such as \fmc{strong} scaling properties and load imbalance.}

We believe that factors such as \alir{the} application representation, scheduling, \alir{the} computing system representation, and system variations, affect the realism of the simulations and deserve further investigation. 
Future work is planned to apply the proposed simulation \fk{approach} to large and \mbox{well-known} performance benchmarks, such as the \fmc{NAS suite, the SPEC suites, the  \mbox{RODINIA} suite,} and other scientific applications.
The development of a tool to automatically transform \alir{the} native application code \alir{into a} simulative one is also envisioned in the future.

\section*{Acknowledgment}
This work has been in part supported by the Swiss National Science Foundation in the context of the ``Multi-level Scheduling in Large Scale High Performance Computers'' (MLS) grant, number 169123 and by the Swiss Platform for Advanced Scientific Computing (PASC) project SPH-EXA: Optimizing Smooth Particle Hydrodynamics for Exascale Computing.

%\section*{References}
\bibliographystyle{ieeetr}
\bibliography{citedatabase}

\end{document}